\documentclass[a4paper,11pt]{article}
\pdfoutput=1 

\usepackage{jheppub} 

\usepackage{amsmath,amssymb,bm,bbm}
\usepackage[utf8]{inputenc}

\title{Correlator correspondences for Gaiotto-Rap\v{c}\'ak dualities and first order formulation of coset models}

\author[a]{Thomas Creutzig}
\author[b]{Yasuaki Hikida}

\affiliation[a]{Department of Mathematical and Statistical Sciences, University of Alberta, Edmonton, \\Alberta T6G 2G1, Canada}
\affiliation[b]{Center for Gravitational Physics, Yukawa Institute for Theoretical Physics, Kyoto University, \\Kyoto 606-8502, Japan}

\emailAdd{creutzig@ualberta.ca}
\emailAdd{yhikida@yukawa.kyoto-u.ac.jp}

\abstract{We derive correspondences of correlation functions among dual conformal field theories in two dimensions by developing a ``first order formulation'' of coset models. We examine several examples, and the most fundamental one may be a conjectural equivalence between a coset $(SL(n)_k \otimes SL(n)_{-1})/SL(n)_{k-1}$ and $\mathfrak{sl}(n)$ Toda field theory with generic level $k$. Among others, we also complete the derivation of higher rank FZZ-duality involving a coset $SL(n+1)_k  /(SL(n)_{k} \otimes U(1))$, which could be done only for $n=2,3$ in our previous paper.  One obstacle in the previous work was our poor understanding of a first order formulation of coset models. In this paper, we establish such a formulation using the BRST formalism. With our better understanding, we successfully derive correlator correspondences of dual models including the examples mentioned above. The dualities may be regarded as conformal field theory realizations of some of the Gaiotto-Rap\v{c}\'ak dualities of corner vertex operator algebras. }

\arxivnumber{2109.03403}
\preprint{YITP-21-91}

\begin{document}
	\maketitle
	\flushbottom


\section{Introduction}

In this paper, we examine dualities in two dimensional conformal field theory admitting extended symmetry with higher spin currents, i.e., W-algebra symmetry. These W-algebras play important roles in recent theoretical physics. For examples,  subsectors of four dimensional gauge theories are known to be organized by W-algebras \cite{Alday:2009aq,Wyllard:2009hg}, and three dimensional higher spin gravity is supposed to be holographic dual to W$_n$ minimal model \cite{Gaberdiel:2010pz}. 

One of the aims of this paper is to derive correspondences of correlation functions of primary operators between conjectural dual theories. Combined with the match of symmetry algebra, we can thus show equivalences of conformal field theories.
The most fundamental example may be the conjectural equivalence between the coset model%
\footnote{The coset may be regarded as an analytic continuation of $(SU(n)_{-k} \otimes SU(n)_{1})/SU(n)_{-k+1}$ with positive integer $-k$. In particular, $SL(n)_{-1}$ is described simply by $n$ complex fermions.}
\begin{align}
 \frac{SL(n)_k \otimes SL(n)_{-1}}{SL(n)_{k-1}} \label{Wcoset}
\end{align}
and $\mathfrak{sl}(n)$ Toda field theory with generic $k$. We note that we use the standard conformal field theory convention for the level; this convention differs by a minus sign from the one often used in mathematics. 
The duality may be regarded as an analytic continuation of the coset realization of W$_n$ minimal model. The coset realization was believed to be true for a long time but it was proven only rather recently \cite{Arakawa:2018iyk}. 
Another famous example of (strong/weak) duality may be the Fateev-Zamolodchikov-Zamolodchikov (FZZ-)duality \cite{FZZ} between $SL(2)_k/U(1)$ coset describing two dimensional cigar model \cite{Witten:1991yr} and sine-Liouville theory proven in \cite{Hikida:2008pe}. Among others, it was applied to a holographic duality in \cite{Kazakov:2000pm}.
In a previous work, we examined extended FZZ-duality involving higher rank coset \cite{Creutzig:2020cmn}
\begin{align}
 \frac{SL(n+1)_k }{SL(n)_{k} \otimes U(1)} \label{hrcoset}
\end{align}
and derived correlator correspondences for $n=2,3$. 
We consider the generalized duality to be important since the coset appears as a dual of higher spin (super-)gravity \cite{Gaberdiel:2010pz,Creutzig:2011fe}. 
In this paper, we complete the derivation for generic $n$. 
We also examine other closely related dualities mentioned below.

In \cite{Hikida:2008pe}, a proof of the original FZZ-duality was given by utilizing the reduction method of $\mathfrak{sl}(2)$ Wess-Zumino-Novikov-Witten (WZNW) model to Liouville field theory \cite{Ribault:2005wp,Ribault:2005ms,Hikida:2007tq}. In order to apply the reduction method, it is important to realize the coset $SL(2)/U(1)$ in terms of product theory $SL(2) \times U(1) \times$ (BRST ghosts) in the BRST formalism  \cite{Gawedzki:1988nj,Karabali:1988au,Karabali:1989dk} and express the $SL(2)$ part in a first order formulation. 
This construction may thus be viewed  as a ``first order formulation'' of a coset model. In our previous paper \cite{Creutzig:2020cmn}, we generalized the analysis by using an extended reduction method from $\mathfrak{sl}(n)$ WZNW model developed in \cite{Creutzig:2015hla,Creutzig:2020ffn}, see also \cite{Hikida:2007sz,Creutzig:2011qm} for previous works.
For a first order formulation of coset model \eqref{hrcoset}, we utilized the analysis of \cite{Gerasimov:1989mz,Kuwahara:1989xy}. However, only the idea and some explicit examples were given in the literature. This is one of the reasons why we could only derive correlator correspondences for $n=2,3$. 
The main idea of  \cite{Gerasimov:1989mz,Kuwahara:1989xy} is to express both the denominator and numerator algebras in terms of Wakimoto free field realizations with free bosons and $(\beta,\gamma)$-systems \cite{Wakimoto:1986gf}. It is not difficult to deal with Cartan directions described by free bosons. However, the parts involving $(\beta,\gamma)$-systems are essentially non-abelian, and it is the part difficult to deal with. The proposal is that some $(\beta,\gamma)$-systems from the numerator algebra cancel with those of the denominator algebra. Indeed, we can see that the central charge of energy momentum tensor matches with the original coset model. However, it was just a conjecture that the coset algebra obtained in this way is isomorphic to the one obtained from the usual Goddard-Kent-Olive (GKO) construction \cite{Goddard:1986ee}. In particular, we do not know how to obtain the interaction terms (or screening operators). Another aim of this paper is to establish the first order formulation of coset model by utilizing the BRST formulation  \cite{Gawedzki:1988nj,Karabali:1988au,Karabali:1989dk} and Kugo-Ogima method \cite{Kugo:1979gm}. We reproduce the cancellation mechanism among $(\beta,\gamma)$-systems and provide the method to give correct interaction terms. In particular, the equivalence to the GKO construction is kept in a manifest way.

The method developed above is not only useful to derive the correlator correspondences with the coset \eqref{hrcoset} of generic $n$ but also strong enough to derive other correlator correspondences like the fundamental case with \eqref{Wcoset}. We also examine the correlator correspondences for the dualities between the coset
\begin{align}
 \frac{SL(n)_k \otimes SL(n)_{1}}{SL(n)_{k+1}} \label{slncoset}
\end{align}
and a theory with a $\mathfrak{gl}(n|n)$-structure, see, e.g., \cite{Litvinov:2016mgi}.  In \cite{Creutzig:2021cyl} a different series of generalized FZZ-duality involving  a Heisenberg coset of a theory with subregular W-algebra symmetry was also derived. That type of generalized FZZ-duality is also called Feigin-Semikhatov duality \cite{Feigin:2004wb} and it is studied from the vertex operator algebra (VOA) perspective in \cite{Creutzig:2020vbt, Creutzig:2021bmz}.
The VOAs serve as symmetry algebras of the conformal field theories and 
the matchings of symmetry algebras for these dual theories are realized as dualities of VOAs.  These VOA dualities were conjectured by Gaiotto and Rap\v{c}\'ak via brane junction realization of VOAs \cite{Gaiotto:2017euk}. The VOAs are denoted as $Y_{n_1,n_2,n_3}[\psi]$ if it is realized as a corner of interfaces between four dimensional gauge theories with gauge group $U(n_1)$, $U(n_2)$ and $U(n_3)$. 
The parameter $\psi$ is related to a coupling constant of gauge theories.
The VOAs are subject to duality relations like 
\begin{align}
Y_{n_1 , n_2 , n_3} [\psi] \simeq Y_{ n_2 , n_3 , n_1} \left[\tfrac{1}{1-\psi}\right] \simeq Y_{n_3,n_1,n_2} \left[1 - \tfrac{1}{\psi} \right] \, .
\end{align}
These Gaiotto-Rap\v{c}\'ak conjectures are a Theorem if at least one of the three labels is zero \cite{Creutzig:2020zaj}.
Among the dualities, a strong/weak one is realized by $Y_{n_1 , n_2 , n_3} [\psi] \simeq Y_{ n_2 , n_1 , n_3} [\psi^{-1}]$.
The equivalence between the coset \eqref{Wcoset} and $\mathfrak{sl}(n)$ Toda field theory corresponds to $Y_{n,0,0} \simeq Y_{0,0,n}$. Moreover, those involving \eqref{hrcoset} and \eqref{slncoset} are related to $Y_{0,n,n+1} \simeq Y_{n,0,n+1}$ and $Y_{0,n,n} \simeq Y_{n,0,n}$, respectively. We remark that the duality analyzed in \cite{Creutzig:2021cyl} is related to $Y_{0,1,n} \simeq Y_{1,0,n}$. 
In this way, we realize  some of  Gaiotto-Rap\v{c}\'ak dualities among VOAs in terms of two dimensional conformal field theory. 

We also derive correlator correspondences involving additional fermionic fields.
Explicitly, we examine another fundamental duality between the coset
\begin{align}
 \frac{SL(2)_k \otimes SL(2)_{-2}}{SL(2)_{k-2}} \label{scoset}
\end{align}
and $\mathcal{N}=1$ super Liouville theory with generic $k$. This can be regarded as an analytic continuation of the coset realization of $\mathcal{N}=1$ minimal model by \cite{Goddard:1986ee}. 
Note that the coset has supersymmetry since $SL(2)_{-2}$ is realized by three free fermions.
We also introduce additional fermions to the coset  \eqref{hrcoset} as
\begin{align}
\frac{SL(n+1)_k \otimes SO(2n)_1}{ SL(n)_{k-1} \otimes U(1)} \, . \label{KScoset}
\end{align}
The coset is a Kazama-Suzuki model \cite{Kazama:1988uz,Kazama:1988qp} with ${ \mathcal N}=2$ superconformal symmetry, and it was proposed to be dual to a ${\mathcal N}=2$ higher spin supergravity \cite{Creutzig:2011fe}.  It was conjectured in \cite{Ito:1990ac,Ito:1991wb} that the coset is dual to $\mathfrak{sl}(n|n+1)$ Toda field theory.
The simplest case with $n=1$ was proven as a mirror symmetry \cite{Hori:2001ax} and in a way similar to the original FZZ-duality in \cite{Creutzig:2010bt}.
The fermionic FZZ duality was utilized to examine  singular Calabi-Yau geometry \cite{Ooguri:1995wj} or its dual picture of NS5-branes \cite{Hori:2002cd}.
We also analyze the coset  \eqref{slncoset} with additional fermions.

\subsection{Organization of the paper}

The paper is organized as follows.
In the next section, we start by reviewing symmetry algebras and the BRST formulation of coset models in order to prepare for later sections.
In particular, we explain the analysis of \cite{Hwang:1993nc}, which shows the equivalence to the GKO construction of coset model.
In section \ref{sec:toda}, we develop a first order formulation of coset models and derive correlator correspondences between the coset \eqref{Wcoset} and $\mathfrak{sl}(n)$ Toda field theory. In subsection \ref{sec:Liouville}, we illustrate our strategy to construct the first order formulation of coset model and derive correlator correspondences for the simplest but non-trivial example with $n=2$. In subsection \ref{sec:hrg}, we then generalize the analysis to the cases with generic $n$. In subsection \ref{sec:N=1}, we consider $\mathcal{N}=1$ supersymmetric case but with $n=2$. In section \ref{sec:HR}, we derive correlator correspondences for higher rank FZZ-duality with generic $n$ by utilizing the first order formulation developed in section \ref{sec:toda}. In section \ref{sec:sHR} we generalize the analysis to the case with $\mathcal{N}=2$ superconformal symmetry. Section \ref{sec:conclusion} is devoted to conclusion and discussions. In appendix \ref{sec:BP}, we  derive the map among different free field realizations of Bershadsky-Polyakov algebra \cite{Polyakov:1989dm,Bershadsky:1990bg} found in \cite{Genra1,Genra2} as field redefinition. This map was a crucial point of the extended reduction method in \cite{Creutzig:2020ffn,Creutzig:2020cmn}.
In appendix \ref{sec:slnndualtiy}, we analyze correlator correspondences between the coset \eqref{slncoset} and a theory with a $\mathfrak{gl}(n|n)$-structure and the case with additional fermions. The analysis is almost the same as in the cases of higher rank FZZ duality and its supersymmetric generalization.

\section{$G/H$ cosets}

In this section we review symmetry and BRST-formulation of $G/H$ cosets.

\subsection{Symmetry algebras of $G/H$ cosets}

In general, it is a difficult problem to precisely determine the full symmetry algebra of a given coset. If $\mathfrak{g}_k$ is the symmetry algebra of a WZNW theory for the Lie group $G$ and $\mathfrak{h}_k$ is the subalgebra corresponding to the subgroup $H$, then the chiral algebra of the coset $G/H$ theory at level $k$ is the subalgebra of $\mathfrak{g}_k$ commuting with the action of $\mathfrak{h}_k$. 
A chiral algebra is called of type $(h_1, \dots, h_s)$ if it is strongly generated by $s$-fields of conformal weight $h_1, \dots, h_s$ and of course one requires that these fields are a minimal generating set. Strongly generated means that every field of the chiral algebra is a normally ordered polynomial in the generating ones and their iterated derivatives. A general theory of relating the type of coset chiral algebras to orbifolds of free theories has been developed in \cite{Creutzig:2014lsa}. 

For example the coset \eqref{scoset} has the same strong generating type as the $SL(2)$-orbifold of three free fermions (in the adjoint representation of $SL(2))$. The latter is of type $(3/2, 2)$ with the spin $3/2$-field being fermionic and the spin two field being the energy-momentum field. 
It is easy to show that the only superalgebra of this type and at central charge $c$ is the $\mathcal N=1$ superconformal algebra at central charge $c$, and this is in fact the simplest case of a uniqueness Theorem of minimal W-superalgebras \cite{Arakawa:2016oyq}. 

For more general cosets it is much more difficult to determine the chiral algebra and we now recall results that are relevant to us. 
Most importantly the GKO-coset of type $SL(n)$ \eqref{Wcoset} has indeed the W$_n$-algebra, that is the principal W-algebra of $\mathfrak{sl}(n)$, as symmetry algebra \cite{Arakawa:2018iyk}. 

The coset algebra of \eqref{hrcoset} coincides with a coset of a W-superalgebra of $\mathfrak{sl}(n+1|n)$. The W-superalgebra has bosonic fields of spin $2, 3, \dots, n+1$ together with a $\mathfrak{gl}(n)$ chiral algebra and $2n$ fermionic fields of conformal weight $(n+2)/2$ and the coset by the $\mathfrak{gl}(n)$ algebra has same symmetry algebra as \eqref{hrcoset}. The levels shifted by the respective dual Coxeter numbers are inverse to each other and so this is a strong/weak duality. This coset duality is a special case of the Gaiotto-Rap\v{c}\'ak triality conjecture \cite{Gaiotto:2017euk} and is proven in \cite{Creutzig:2020zaj}.
The coset \eqref{slncoset} is closely related to this one and the symmetry algebra coincides in this case with the chiral algebra of a $\mathfrak{gl}(n)$-coset of a W-superalgebra of $\mathfrak{gl}(n|n)$. Again the critically shifted levels of the two theories are inverses of each other; moreover this duality is also contained in the Gaiotto-Rap\v{c}\'ak  triality conjecture \cite{Gaiotto:2017euk} that is proven in \cite{Creutzig:2020zaj}. 
For the supersymmetric version one currently only knows that the symmetry algebra of the coset has the same type as the principal W-superalgebra of $\mathfrak{gl}(n|n)$ \cite{Creutzig:2012sf}; a full proof is in addition under current investigation. 
The situation of the Kazama-Suzuki coset \eqref{KScoset} is similar. It is known \cite{Creutzig:2014lsa} that the type is indeed $(1, 3/2, 3/2, 2, 2, 5/2, 5/2, 3, 3, \dots, n+1/2, n+1.2, n+1)$, that is the same as the principal W-superalgebra of $\mathfrak{sl}(n+1|n)$. A proof that algebras coincide is however only known for $n=1, 2$ \cite{Genra:2019tgw}.

\subsection{BRST formulation of $G/H$ cosets}

Applying previous works \cite{Hikida:2008pe,Creutzig:2010bt,Creutzig:2020cmn,Creutzig:2021cyl}, we relate correlation functions of coset models to those of dual theories. For this, it is useful to express a coset model $G/H$ by a gauged WZNW model in the BRST formulation developed in \cite{Gawedzki:1988nj,Karabali:1988au,Karabali:1989dk}.
In the formulation, the coset model is described by a product theory $G \times H \times$ (BRST ghosts), and physical states are obtained as elements of the cohomology of BRST charge. 
In this section, we collect useful results about the BRST formulation of $G/H$ coset models to prepare for later sections. In particular, we outline the analysis on the equivalence between the BRST formulation and GKO construction by \cite{Goddard:1986ee} given in \cite{Karabali:1989dk} for abelian $H$ and extended in \cite{Hwang:1993nc} for non-abelian $H$.
This procedure has a mathematical counterpart that is developed by Frenkel, Garland and Zuckerman \cite{Frenkel:1986dg} and is summarized in section 2  of \cite{Creutzig:2019kro}.

For simplicity of explanation, let us assume that $G$ is a simple group, though it is straightforward (and will be used) for arbitrary reductive $G$. We denote the action of the WZNW model at level $k$ by 
$
S^\text{WZNW}_k[g] = \int d^2 z \mathcal{L}^\text{WZNW}_k [g]
$
 with $g \in G$. Then the theory has the symmetry of current algebra generated by $J (z)= J^A ( z) t_A , \bar J (\bar z)= \bar J^A ( \bar z) t_A$, where the $t_A$ form a  basis of Lie algebra $\mathfrak{g}$ of the group $G$. The mode expansions satisfy
\begin{align}
[J^A_m , J^B_n] = i f^{AB}_{~~C} J^C_{m+n} + \frac{k}{2} m \delta_{m,-n}
 g^{AB} \, , 
\end{align}
where $ f^{AB}_{~~C} $ are structure constants of $\mathfrak{g}$ and $g^{AB}$ is a metric on $G$.
We have similar commutation relations  for $\bar J^A_n$.

Now we would like to gauge a subgroup $H$ of $G$. We assume again that $H$ is a simple group and denote the level for $H$ by $k_H$.  In the BRST formulation, the effective action is given by
\begin{align}
S = S^\text{WZNW}_k[g] + S^\text{WZNW}_{- k_H - 2 c_H}[\tilde h] + (\text{BRST ghosts}) \, ,
\end{align}
where $\tilde h \in H$ and $c_H$ is the dual Coxeter number of $H$. Let $I$ be an index set, such that $\{ x^a | a \in I \}$ is a basis of  the Lie algebra $\mathfrak{h}$ of $H$. The BRST ghosts have the same indices as $\mathfrak{h}$ and are denoted by $c_a (z), b^a (z)$. They are Grassmann odd fields and their conformal dimensions are $0,1$, respectively.  Their operator product expansions (OPEs) are
\begin{align}
c_a(z) b^{a'} (w) \sim \frac{\delta_a^{~a'}}{z -w} \, .
\end{align}
In the GKO construction, the physical states are obtained by the condition
\begin{align}
J^a_n |\text{phys} \rangle = 0 \label{GKOphys}
\end{align}
for $n > 0$. Here $J^a_n$ are generators of the subsector $\mathfrak{ h }_{k_H } \subset \mathfrak{ g }_k$ and in particular the level is given by $k_H$.
On the other hand, in the BRST formulation, there are additional currents $\tilde J^a$ with level $k_{\tilde H} = - k_H - 2 c_H$ from the action  $S^\text{WZNW}_{- k_H - 2 c_H}[\tilde h]$.
Using these two types of currents, the BRST charge for the holomorphic part can be defined as
\begin{align}
Q = \oint \frac{dz}{ 2 \pi i} \left[ c_a (z)(J^a (z) + \tilde J^a (z)) - \frac{i}{2} f^{ad}_{~~e } c_a (z) c_d (z) b^e (z) \right] \, . \label{QBRST}
\end{align}
Here and in the followings,  the normal ordering prescription is assumed for any product of fields and Einstein's summation convention is in place.
The physical state condition is then expressed as
\begin{align}
Q |\text{phys} \rangle = 0 \, . \label{Qphys}
\end{align}
In fact since $Q$ is nilpotent, $Q^2 = 0$, the physical states are obtained as the elements of $Q$-cohomology.

The energy momentum tensor is given by Sugawara construction as
\begin{align}
\begin{aligned}
T(z) & = \frac{1}{k+c_G} J_A(z) J^A (z) - \frac{1}{k _H + c_H} \tilde J_a(z) \tilde J^a (z) - b^a (z) \partial c_a  (z) \\
& \equiv T^G (z)+ \tilde T^H (z)+ T^\text{gh} (z)\, .
\end{aligned}
\end{align}
Note that the energy momentum tensor in the GKO construction is  $T^\text{GKO} = T^G - T^H$.
It was shown in \cite{Karabali:1989dk} that the above one can be written as  $T = T^\text{GKO} + T^\text{tot}$, where $T^\text{tot}$ was shown to be BRST exact as 
\begin{align}
T^\text{tot} (z)= T^H (z)+ \tilde T^H (z)+ T^\text{gh} (z) = \frac{1}{k_H + c_H} [Q , b_a (z) (J^a (z) - \tilde J^a (z))] \, .
\label{Ttot}
\end{align}
In particular, the total energy momentum tensor has zero central charge.
We may also define total currents by
\begin{align}
J^{\text{tot},a} (z) = [ Q , b^a (z)] = J^a (z)+ \tilde J^a (z)+ J^{\text{gh},a} (z) \, . \label{Jtot}
\end{align}
We can check that the currents generate the affine Lie algebra $\mathfrak{ h}_0$ at level zero. Here we have defined $\mathfrak{h}_{2c_H}$ currents at level $2c_H$ by $J^{\text{gh},a} = i f^{ad}_{~~e} b^e (z)c_d (z)$.

The equivalence between the GKO construction and the BRST formulation was shown in \cite{Karabali:1989dk} for abelian $H$ and it was extended in \cite{Hwang:1993nc} for non-abelian $H$ as mentioned above.
In the following, we explain the result of  \cite{Hwang:1993nc}.
We would like to study the solutions to the physical condition \eqref{Qphys}.
Generic states may be constructed from primary states $|R_G, R_{\tilde H} , 0 \rangle$ satisfying
\begin{align}
J^A_n |R_G, R_{\tilde H} , 0 \rangle = \tilde J^a_n |R_G, R_{\tilde H} , 0 \rangle = 0
\end{align}
for $n>0$. The labels $R_G,R_{\tilde H}$ denote the representations of $G, \tilde H$, respectively.
For the ghost sector, we assign
\begin{align}
c^a_n |R_G, R_{\tilde H} , 0 \rangle = 0 \quad (n \geq 1) \, , \quad b^a_n |R_G, R_{\tilde H} , 0 \rangle = 0 \quad (n \geq 0) \, .
\end{align}
Generic states $|s\rangle$ are constructed by acting with negative modes on the primary states.

Consider the eigenstates $|s\rangle$ of $J^{\text{tot},i}_0$ of eigenvalue (or weight) $ \mu^{\text{tot},i}$, that is
\begin{align}
J^{\text{tot},i}_0|s\rangle = \mu^{\text{tot},i} |s\rangle \, .
\end{align}
Here $i$ labels the Cartan subalgebra of $\mathfrak{h}$ and runs over $i=1,2,\ldots,r_\mathfrak{h}$ with $r_\mathfrak{h}$ the rank of $\mathfrak{h}$.
We assume that the physical state condition $Q |s \rangle = 0$  is satisfied. If $\mu^{\text{tot},i} \neq 0$, then we can write
\begin{align}
	| s \rangle = \frac{1}{\mu^{\text{tot},i}} [ Q , b^i_0] 	| s \rangle = \frac{1}{\mu^{\text{tot},i}} Q b_0^i 	| s \rangle \, . \label{trivial}
\end{align}
Therefore, non-trivial elements in the cohomology have to come from the sector with zero weights, that is $\mu^{\text{tot},i}=0$ for all  $i=1,2,\ldots,r_\mathfrak{h}$.

Instead of considering the BRST charge $Q$ defined in \eqref{QBRST}, we consider the cohomology of $\hat Q$ defined as  \cite{Hwang:1993nc}%
\footnote{In \cite{Hwang:1993nc} the BRST charge $Q$ is decomposed as in \eqref{BRSTdec} in order to show the equivalence between the GKO construction and the BRST formulation. The Hilbert space is actually doubled in the BRST formulation due to the degeneracy of vacua in ghost system, which can be avoided by assigning the extra condition of \eqref{b0cod}, see \cite{Hwang:1993nc} for more details.
We shall implicitly use the result by choosing physical states of the GKO construction as non-trivial elements of the BRST cohomology.}
\begin{align}
Q = \hat Q + M_i b_0^i + c_{0,i} J^{\text{tot},i}_0 \, . \label{BRSTdec}
\end{align}
As argued above, we restrict ourselves to the sector with  zero eigenvalue states of $J^{\text{tot},i}_0$ as
\begin{align}
J^{\text{tot},i}_0 |s \rangle = 0 \label{J0cond}
\end{align}
for all $i=1,2,\ldots,r_\mathfrak{h}$. 
The operator $\hat Q$ is nilpotent if its action is restricted to this sector. We also study the cohomology of $\hat Q$ on the relative space
\begin{align}
b_0^i |s \rangle = 0 \label{b0cod}
\end{align}
for all $i=1,2,\ldots,r_\mathfrak{h}$. Then, the non-trivial elements of the relative cohomology  are given by states $|\phi \rangle$ with the following properties  \cite{Hwang:1993nc}. They have zero ghost number and no action of $\tilde J^a_{-n}$ with $n > 0$. Moreover, they satisfies
\begin{align}
J^a_n | \phi \rangle = b_n^a | \phi \rangle = c_n^a |\phi \rangle = 0
\end{align}
for $n > 0$. This is the same as the physical condition in the GKO construction \eqref{GKOphys} along with the decoupling of other sectors.

We may study some implications of the analysis.
Firstly, we observe that  the condition  $J^\text{tot,i}_0 |\phi \rangle = 0$ implies that
\begin{align}
\mu^i + \tilde \mu^i  + 2  \rho^i = 0 \, . \label{Jcond}
\end{align}
Here $\rho$ is the Weyl vector defined as usual
\begin{align}
\rho = \frac12 \sum_{\alpha \in \Delta_+} \alpha \, , \label{Weyl}
\end{align}
where the sum is over all positive root $\alpha \in \Delta_+$. Moreover, $\mu^i , \tilde \mu^i$ are weights with respect to $\mathfrak{h}$. We consider
\begin{align}
L^\text{tot}_0 |\text{phys} \rangle = \left[ \frac{1}{k_H+c_H} (C (h)- C (\tilde h)) + N_J + N_{\tilde  J} + N_{\text{gh}}\right] |\text{phys} \rangle  \, .
\end{align}
Here $N_J,N_{\tilde J},N_\text{gh}$ are the number operators of $J^a_n,\tilde J^a_n$ and BRST ghosts, respectively, and they are zero for physical states as mentioned above. Here, $C (h), C(\tilde h)$ are the second-order Casimir operators of the Lie algebras generated by $J^a_0$ and $\tilde J^a_0$, respectively. As argued around \eqref{trivial}, non-trivial elements of cohomology are given by eigenstates with zero eigenvalues.
Thus we have to set
\begin{align}
C (h)=  C (\tilde h) \, . \label{Ccond}
\end{align}

\section{Toda field theories form coset models}
\label{sec:toda}

In the previous section, we reviewed generic properties of BRST formulation of $G/H$ cosets.
One of the aims of this paper is to establish the conjectured method of  \cite{Gerasimov:1989mz,Kuwahara:1989xy}.
For this, we adopt a first order formulation for $G$ and $H$ WZNW models and observe cancellations among $(\beta,\gamma)$-fields.  In this section, we establish the method using the BRST formulation and apply it to the coset of the type \eqref{Wcoset}.
In order to illustrate our procedures, we first examine the simplest but non-trivial example,
\begin{align}
	\frac{SL(2)_k \otimes SL(2)_{-1}}{SL(2)_{k-1}} \label{Vircoset}
\end{align}
and relate it to Liouville field theory.
In subsection \ref{sec:hrg}, we generalize the analysis to the coset \eqref{Wcoset} with generic $n$, and in subsection \ref{sec:N=1} we introduce $\mathcal{N}=1$ supersymmetry to the case with \eqref{Vircoset}.

\subsection{Liouville field theory from coset model}
\label{sec:Liouville}

Applying the BRST formulation reviewed in the previous section,  the action for the coset \eqref{Vircoset} consists of four parts as
\begin{align}
	S = S_k^\text{WZNW} [\phi, \beta , \gamma ] + S_\psi [\psi] + S_{- k + 5}^\text{WZNW} [\tilde \phi , \tilde \beta , \tilde \gamma] + S_{bc} [b^a,c_a ] \, .  \label{BRSTaction}
\end{align}
The action is invariant under BRST transformation and physical states are obtained as non-trivial elements of BRST cohomology.

The first and  third actions describe $\mathfrak{sl}(2)$ WZNW models at the level $k$ and $-k + 5$, respectively.
An important point here is to use the first order formulation of the $\mathfrak{sl}(2)$ WZNW model at level $k$.
The action is  given by
\begin{align}
	S_k^\text{WZNW} [\phi , \gamma , \beta] = \frac{1}{2 \pi} \int d^2 w 
	\left ( \partial \phi \bar \partial \phi - \beta \bar \partial \gamma - \bar \beta \partial \bar \gamma + \frac{b}{4} \sqrt{g} \mathcal{R} \phi  + \lambda \beta \bar \beta e^{2 b \phi}\right) \label{WZNWaction}
\end{align}
with $b = 1/\sqrt{k-2}$.
The conformal weights of $(\beta , \gamma)$ are $(1,0)$.
Here $g_{\mu\nu}$ is the background metric and $g =\det g_{\mu \nu}$. Moreover, $\mathcal{R}$ is the curvature of the worldsheet. We take the metric as $d s^2 = |\rho (z)|^2 dz d\bar z $ with $\rho (z) =1$ for almost all the cases. The symmetry is the $\mathfrak{sl}(2)$ current algebra generated by
\begin{align}
J^+  = \beta  \, , \quad J^3  = b^{-1} \partial \phi - \beta  \gamma  \, , \quad 
J^- = \beta \gamma \gamma - 2 b^{-1} \gamma \partial \phi + k \partial \gamma \, .
\end{align}
For the third summand of the action in \eqref{BRSTaction}, we denote the generators of the current algebra by $\tilde J^a$ with $a=\pm,3$.

The second factor in the numerator of \eqref{Vircoset} corresponds to a pair of free fermions. The action may be given by
\begin{align}
	S_{\psi}[\psi] = \frac{1}{2 \pi} \int d^2 w  \left[  \psi^+ \bar \partial \psi^- +  \bar \psi^+ \partial \bar \psi^- \right] \, , \label{Spsi}
\end{align}
where the conformal weights of $(\psi^+ ,  \psi^-)$ are $(1/2,1/2)$.
It is convenient to bosonize the free fermions by
\begin{align}
	\psi^\pm = e^{\pm i \sqrt{2} H^L} \, , \quad H^L (z) H^L (0) \sim-  \frac12 \ln z 
\end{align}
and similarly for $\bar \psi^\pm$ written by $H^R$. We further define $H = H^L + H^R$.
The $\mathfrak{sl}(2)$ current generators are given by
\begin{align}
	J^+_\psi = - e^{ 2 i  H^L} \, , \quad J^3_\psi = i \partial H^L \, , \quad J^-_\psi = e^{- 2 i  H^L} \, .
\end{align}

The final term of the action is for the BRST ghosts, and in the current case it can be written as
\begin{align}
	S_{bc}[b^a , c_a] = \frac{1}{2 \pi} \int d^2 w \sum_{a =  \pm , 3} \left[  b^a \bar \partial c_a +  \bar b^a \partial \bar c_a \right] \, ,
\end{align}
where the conformal weights of $(b^a,c_a)$ are $(1,0)$.  We may define $\mathfrak{sl}(2)$ current generators as
\begin{align}
	J^+_{bc} = - \sqrt{2} (b^+ c_3 +  b^3 c_- ) \, , \quad J^3_{bc} =  b^+ c_+ - b^- c_- \, , \quad J^-_{bc} =  \sqrt{2} (b^- c_3 + b^3 c_+ ) \,  .
\end{align}
With the BRST ghosts, the BRST charge \eqref{QBRST} can be written as
\begin{align}
Q = \oint \frac{dz}{2 \pi i}  \left[ c_a (z) \left(J^a (z)+ J^a_\psi (z) + \tilde J^a (z) + \frac{1}{2} J_{bc}^a (z) \right) \right] \, . \label{BRSTcharge}
\end{align}
One can explicitly verify that  $Q^2 = 0$.

\subsubsection{Primary states}

We examine correlation functions of the coset model in the BRST formulation. 
In particular, we  show that every $N$-point function of the coset \eqref{Vircoset} can be mapped to an $N$-point function of Liouville field theory. It is well known that the symmetry of  both side agrees with each other, and hence we just need to examine correlation functions of primary operators. In this section, we examine primary states in the coset model \eqref{Vircoset} and reduce it to those of Liouville field theory.
We follow the arguments in the section 4 of \cite{Polchinski:1998rq}, which is based on the analysis of \cite{Kugo:1979gm}.
We consider primary operators in the BRST formulation of the coset  \eqref{Vircoset}.
For this, it is enough to consider the vertex operators of the form
\begin{align}
	V = \mathcal{P} (\gamma , \tilde \gamma) e^{2 b j \phi} e^{ 2  i s H } e^{2 \tilde b \tilde\jmath \tilde \phi } \label{op}
\end{align}
with an arbitrary function $\mathcal{P} $ of $\gamma , \tilde \gamma$ (and anti-holomorphic counterparts) and $\tilde b = 1/\sqrt{-k+3}$. In the following, we show that the same correlation function can be obtained by projecting $\beta, \gamma$ and $\tilde \beta , \tilde \gamma$, see \cite{Bershadsky:1989mf} for a similar analysis.

We first define an operator
\begin{align}
	N^{\beta \gamma} = - \sum_{m= - \infty }^\infty \beta_{-m} \gamma_m \, , \quad 	[\beta_m , \gamma_n] = \delta_{m,-n} \, , \label{Ngh}
\end{align}
which counts the number of $\beta$ minus the number of $\gamma$. With this, we can decompose the BRST generator as
\begin{align}
	Q = Q_1 + Q_0 + Q_{-1} \, , 
\end{align}
where the subscript represents the eigenvalue of $N^{\beta \gamma}$. In particular, 
we have
\begin{align}
	Q_1 = \sum_{m=-\infty}^\infty \beta_{-m} c_{+,m} \, ,
\end{align}
which satisfies $(Q_1)^2 = 0$. We further define
\begin{align}
	R = \sum_{m = - \infty}^\infty \gamma_{-m} b^+_{m} \, ,
\end{align}
which leads to
\begin{align}
	S = \{ Q_1 , R\} = \sum_{m=1}^\infty \beta_{-m} \gamma_m + \sum_{m=0}^\infty \gamma_{-m} \beta_{m} 
	 - \sum_{m=1}^\infty b^+_{-m} c_{+ ,m} + \sum_{m=0}^\infty c_{+,-m} b^+_{m} \, . \label{S}
\end{align}
The eigenvalue of $S$ is not necessary a non-negative number, but this does not cause any problems for the restricted form of vertex operators as in \eqref{op}. For the operators of the form \eqref{op}, $S$ simply counts the number of $\gamma$ in the function $\mathcal{P}$. Since $S$ commutes with $Q_1$, we can consider the eigenfunction of $S$ in the cohomology  of $Q_1$ as $S|\phi \rangle = s |\phi  \rangle $. 
As argued around \eqref{trivial}, non-trivial elements in the cohomology of $Q_1$ should come from the sector with $s=0$, i.e., without $\gamma$-dependence in the vertex operators.

Above, we have shown that a non-trivial element of $Q_1$-cohomology does not depend on $\gamma$.
We then map the cohomology of $Q_1$  to that of $Q$.
In order to do so, we introduce another operator
\begin{align}
	U = \{  Q_0 + Q_{-1} , R \} \, .
\end{align}
If $S | \phi \rangle  = 0$, then the state
\begin{align}
	|\phi ' \rangle = (1 - S^{-1} U + S^{-1} U S ^{-1} U - \cdots ) 	|\phi \rangle  \label{Vp}
\end{align}
is also annihilated by $S+U$. As in the argument above,  the only non-trivial elements of $Q$-cohomology come from the states satisfying $(S+U) |\phi ' \rangle  = 0$. 
Thus we can map a non-trivial element of the $Q_1$-cohomology without any BRST ghosts to an element of $Q$-cohomology.%
\footnote{Here we remark that the cohomology of  $Q_1$ is not isomorphic to the cohomology of $Q$ since the operator $S$ defined in \eqref{S} does not always have non-negative eigenvalues. The current analysis is enough if the vertex operators in the BRST formulation of the coset \eqref{Vircoset} do not include any BRST ghosts. However, if we want to analyze the equivalence of Hilbert space of two dual theories, then the current analysis need to be at least modified.
}

\subsubsection{Correlation functions}

Conversely, a vertex operator in  $Q$-cohomology  may be put in the form of \eqref{Vp}. Note that $S^{-1}U$ always decreases the eigenvalue of the operator $N^{\beta , \gamma}$ defined in \eqref{Ngh}.
On the other hand, $\beta$ in the action \eqref{WZNWaction} can be replaced by (see section 9 of \cite{Arakawa:2018iyk} as well)
\begin{align}
	\beta (w)- \oint_w \frac{dz}{2 \pi i}  b^+ (z) Q (w) = e^{2 i H^L} (w) - \tilde \beta (w) + \sqrt{2} (b^+ c_3 (w) + b^3 c_- (w)) \, .  \label{rbeta}
\end{align}
Therefore, the total action  \eqref{BRSTaction} is in the same cohomology class as an action that contains 
no operator increasing the $N^{\beta , \gamma}$-eigenvalue.
With this choice we can safely use vertex operators without $\gamma$.

We can obtain more restrictions on the vertex operators, which come from the physical conditions \eqref{Qphys}.
The total currents  in \eqref{Jtot} become
\begin{align}
	J^\text{tot,3}_m \equiv \{ Q , b^3_m \} = J^3_m + J^3_{\psi,m} + \tilde J^3_m + J_{bc,m}^3
\end{align}
and the total energy momentum tensor in \eqref{Ttot} is
\begin{align}
  L^\text{tot}_m \equiv \{Q , \tfrac{1}{k - 3} \sum_{n} (J^a_{m+n} + J^a_{\psi,m+n} - \tilde J^{a}_{m+n}) b_{-n , a}\} \, .
\end{align}
Considering the vertex operator of the form
\begin{align}
	V = \tilde \gamma^m \bar{\tilde \gamma}^{\bar m}  e^{2 b j \phi} e^{2 i s H } e^{2 \tilde b \tilde\jmath \tilde \phi } \, , 
\end{align}
the conditions corresponding to \eqref{Jcond} and \eqref{Ccond} become
\begin{align}
	- j - m + s  - \tilde \jmath  + 1 = 0 \, , \quad - \frac{(j-s)(j - s -1)}{k-3} + \frac{\tilde \jmath ( \tilde \jmath - 1)}{k - 3} = 0 \, . \label{Virconst}
\end{align}
We have also conditions with $m$ replaced by $\bar m$.
The solution is
\begin{align}
	\tilde \jmath = 1 - j + s \, , \quad m = 0 \, , 
\end{align}
that is, we should use the vertex operator of the form
\begin{align}
	V = e^{2 b j \phi} e^{ 2 i  s H } e^{2 \tilde b (1 - j + s) \tilde \phi } \, .\label{Vnew}
\end{align}
In particular, there is no dependence on $\tilde \gamma$ and BRST ghosts, and hence we can neglect $\tilde \beta$ and BRST ghosts in \eqref{rbeta}.

As in \eqref{Vnew}, there are three fields $\phi,H,\tilde \phi$ involved now. 
The action for them is
\begin{align}
	S [\phi,H,\tilde \phi] = \frac{1}{ 2 \pi} \int d^2 w \left[ \partial \phi \bar \partial \phi + \partial H \bar \partial H +  \partial \tilde \phi \bar \partial \tilde \phi + \frac{\sqrt{g} \mathcal{R} }{4} (b \phi + \tilde b \tilde \phi) 
	+ \lambda e^{2 b \phi + 2 i  H } \right] \, .
\end{align}
Rotating the fields as
\begin{align}
	b \phi + i H = b ' \phi ' \, , \quad - i \phi + b H = b ' H ' \, , \quad b' = \sqrt{\frac{3 -k}{k -2}} \, ,
\end{align}
the  action becomes
\begin{align}
\begin{aligned}
	S [\phi ' ,H ' ,\tilde \phi] &= \frac{1}{ 2 \pi} \int d^2 w \left[ \partial \phi ' \bar \partial \phi ' + \partial H ' \bar \partial H '+  \partial \tilde \phi \bar \partial \tilde \phi \right] \\
 & \quad +
\frac{1}{ 2 \pi} \int d^2 w \left[ \frac{\sqrt{g} \mathcal{R} }{4} \left( Q_{\phi '} \phi ' + Q_{H'} H' + Q_{\tilde \phi } \tilde \phi \right) 
	+ \lambda e^{2 b ' \phi ' } \right]
\end{aligned}
\end{align}
with the background charges
\begin{align}
	Q_{\phi '} = b' + 1/b' \, , \quad Q_{H '} = - i \tilde b \, , \quad Q_{\tilde \phi} = \tilde b \, .
\end{align}
The vertex operator \eqref{Vnew} is now changed as 
\begin{align}
	V = e^{ 2 ((b' + 1/b') j - 1/b ' s) \phi '} e^{ - 2i  \tilde b (j- s) H '} e^{2 \tilde b (1 - j + s) \tilde \phi } \, .
\end{align}
For $N$-point functions, we can see that the contributions from $H'$ and $\tilde \phi $ cancel out.%
\footnote{The cancellation occurs up to the coefficients coming from the use of reflection relation. The same is true for the arguments below.}
In the language of \cite{Gerasimov:1989mz,Kuwahara:1989xy}, the field space spanned by $\phi,H$ is restricted to be orthogonal to that spanned by $H'$.
We thus end up with the Liouville correlation function as
\begin{align}
	\left \langle \prod_{\nu=1} ^N V_\nu (z_\nu) \right \rangle \, , \quad 	V_\nu (z_\nu)= e^{2 ((b' + 1/b') j_\nu - 1/b ' s_\nu) \phi ' (z_\nu)} \label{Lcorr}
\end{align}
with the action
\begin{align}
	S [\phi '  ] = \frac{1}{ 2 \pi} \int d^2 w \left[ \partial \phi ' \bar \partial \phi '  + \frac{\sqrt{g} \mathcal{R} }{4  }(b' +1/b') \phi ' 
	+ \lambda e^{2 b ' \phi ' } \right] \, . \label{Laction}
\end{align}
In this way, we have shown that the computation of correlation functions of the coset \eqref{Vircoset} reduces to that of \eqref{Lcorr} with the action of Liouville field theory in \eqref{Laction}.

\subsection{Higher rank generalization}
\label{sec:hrg}

In this subsection,  we examine the coset \eqref{Wcoset} with generic $n$ and derive $\mathfrak{sl}(n)$ Toda field theory as in the case with $n=2$. 
The action in the BRST formulation is similar to \eqref{BRSTaction} and given by
\begin{align}
	S = S_k^\text{WZNW} + S_\psi + S_{- k  + 1 + 2 n}^\text{WZNW} + S_{bc} \, . 
\end{align}
We explain each part below.

The WZNW model based on $\mathfrak{sl}(n)$ Lie algebra at level $k$ is represented by  $S_k^\text{WZNW} $.
As in the case of $\mathfrak{sl}(2)$, we use the action in the first order formulation as (see, e.g., \cite{Kuwahara:1989xy})
\begin{align}
\begin{aligned}
	S_k^\text{WZNW}& = \frac{1}{2 \pi} \int d^2 w 
	\left [ \frac{G^{(n)}_{ab}}{2}\partial \phi^a \bar \partial \phi^b - \sum_{i > j}^n (  \beta_{i,j} \bar \partial \gamma_{i,j} + \bar \beta_{i,j} \partial \bar \gamma_{i,j} ) +  \frac{b}{4} \sqrt{g}  \mathcal{R} \sum_{a=1}^{n-1} \phi^a \right]  \\
& \quad +  \frac{\lambda }{2 \pi} \int d^2 w 
	\sum_{j=1}^{n-1} V_{j+1 ,j}  \, .
\end{aligned}
\end{align}
Here $G^{(n)}_{ab}$ is the Cartan matrix of $\mathfrak{sl}(n)$ and $a=1,2,\ldots , n-1$. The inverse of the matrix is defined by $G^{(n)ab} G^{(n)}_{bc} = \delta^{a}_{~c}$ and the index is raised as $\phi^a = G^{(n)ab } \phi_b$. Moreover, we set
 $b = 1/\sqrt{k-n}$. The indices $i,j$ run over $i,j=1,2,\ldots , n$. The interaction terms are
 \begin{align}
 V_{j+1 ,j} = \left |\beta_{j+1 , j} + \sum_{l=1}^{j-1}\beta_{j+1, l} \gamma_{j,l} \right |^2  e^{b \phi_j} \label{int}
 \end{align}
with $j=1,2,\ldots, n-1$. 
In the case with $n=2$, the interaction term includes only $\beta$, so we just had to take care of that term. However, in the current case, the interaction terms depend on $\gamma_{i,j}$ as well. We will see below that they do not cause any problems.
The symmetry of this model is $\mathfrak{sl}(n)$ current algebra.
Among the generators, $J_{i,j}$ with $i<j$ are given by
\begin{align}
	J_{i,j} = \beta_{j,i} + \sum_{l = j +1}^n \gamma_{l,j} \beta_{l,i} \, . \label{slJij}
\end{align}
The Cartan direction is generated by
\begin{align}
	H_a = \hat H_{a} - \hat H_{a+1} \label{slHa}
\end{align}
with
\begin{align}
	\hat H_a = b ^{-1} \partial \varphi_a + \sum_{l=1}^{a-1} \gamma_{a,l} \beta_{a,l} - \sum_{l = a+1}^n \gamma_{l,a} \beta_{l,a} \, .   \label{slhatHa}
\end{align}
The free bosons are introduced as $\phi_a = \varphi_{a} - \varphi_{a+1}$ with $\varphi_a (z) \varphi_b (0) \sim - \delta_{a,b} \ln |z|^2$. The other currents $J_{i,j}$ with $i > j$ are fixed so as to reproduce the OPEs.
For the third term, we use $\tilde b = 1/\sqrt{ -k + 1 + n}$ and $\tilde J_{i,j}$, $\tilde H_a$ for $\mathfrak{sl}(n)$ currents.

The second factor in the numerator of  the coset \eqref{Wcoset} is described by $n$ pairs of free fermions.
We may use its action as
\begin{align}
	S_{\psi} = \frac{1}{2 \pi} \int d^2 w  \sum_{j=1}^n \left[  \psi^+_j \bar \partial \psi^-_j +  \bar \psi^+_j \partial \bar \psi^-_j \right] 
\end{align}
with conformal weight $1/2$ for $\psi_j^\pm$ and similarly for $\bar \psi_j^\pm$ .
We may bosonize the free fermions by
\begin{align}
	\psi^\pm_j = e^{\pm i Y^L_j} \, , \quad Y^L_i (z) Y^L_j (0) \sim-  \delta_{i,j} \ln z
\end{align}
and similarly for $\bar \psi^\pm_j$ written by $Y^R_j$. We further define $Y_j = Y^L_j + Y^R_j$.
The $\mathfrak{sl}(n)$ currents are given by
\begin{align}
	J^\psi_{i,j} = - \psi^+_i \psi^-_j   \quad (i < j)\, , \quad H^a = \psi^+_{a} \psi^-_{a} -  \psi^+_{a+1} \psi^-_{a+1}
\end{align}
and similarly for $	J^\psi_{i,j}$ with $i > j$.

The action for BRST ghosts may be written as
\begin{align}
	S_{bc}= \frac{1}{2 \pi} \int d^2 w \left[  \sum_{i \neq j}^n  \left(  b^{i,j} \bar \partial c_{i,j} +  \bar b^{i,j} \partial \bar c_{i,j} \right) + \sum_{a=1}^{n-1} \left( b^a \bar \partial c_a +  \bar b^a \partial \bar c_a \right) \right] \, ,
\end{align}
where the conformal weights of $(b^{i,j},c_{i,j})$ and $(b^a,c_a)$ are $(1,0)$.  The $\mathfrak{sl}(n)$ currents consisting of BRST ghosts are denoted as $J^{bc}_{i,j}$ and $H^{bc}_a$. With these BRST ghosts, the BRST charge \eqref{QBRST} becomes
\begin{align}
	\begin{aligned}
	Q &= \oint \frac{dz}{2 \pi i}  \sum_{ i \neq j}^n \left[ c_{j,i} (z) \left(J_{i,j} (z)+ J^\psi_{i,j} (z) + \tilde J_{i,j} (z) + \frac{1}{2} J^{bc}_{i,j} (z) \right) \right] \\
	&\quad +  \oint \frac{dz}{2 \pi i}  \sum_{ a =1}^{n-1} \left[ c_a (z) \left(H_a (z)+ H^\psi_a (z) + \tilde H_a (z) + \frac{1}{2} H^{bc}_a (z) \right) \right] \, .
    \end{aligned}
\end{align}

As  in the case with $n=2$, we consider the correlation functions of vertex operators of the form
\begin{align}
	V = \mathcal{P}(\gamma_{i,j} , \tilde \gamma_{i,j}) e^{b j \cdot \varphi} e^{ i s \cdot Y} e^{ \tilde  b \tilde \jmath \cdot \tilde \varphi} \, . \label{WV}
\end{align}
In order for $\sum_{l=1}^n \varphi_l$ and $\sum_{l=1}^n \tilde \varphi_l$ decouple, $\sum_{l=1}^n j^l = \sum_{l=1}^n \tilde \jmath^l = 0$ have to be satisfied.
Moreover, we decompose the BRST charge $Q =Q_1 + Q_0 + Q_{-1}$ by the eigenvalues of the operator
\begin{align}
	N^{\beta \gamma}  = - \sum_{i<j}^n \sum_{m=- \infty}^\infty \beta_{j,i,-m} \gamma_{j,i,m} \, . \label{WN}
\end{align}
As shown above, we can construct a map between the elements of $Q_1$-cohomology and those of $Q$-cohomology.
Moreover, we replace $\beta_{j,i}$ in the interaction term \eqref{int} by
\begin{align}
	\beta_{j,i} (w) - \oint \frac{dz}{2 \pi i} b^+ _{i,j} (z)Q (w) = e^{i (Y_i - Y_j) }(w) - \tilde \beta_{j,i}(w) - J^{bc}_{i,j} (w) \, . \label{betaij}
\end{align}
A difference from the case with $n=2$ is the dependence of $\gamma_{i,j}$ in the interaction terms \eqref{int}.
However, they only decrease the eigenvalue of $N^{\beta \gamma}$ in \eqref{WN}.
Removing all $\beta_{j,i}$ by applying \eqref{betaij}, there are no fields raising  the eigenvalue of $N^{\beta \gamma}$. Thus we can safely neglect all the terms including  $\gamma_{i,j}$  in the interaction terms along with the vertex operators of the form  \eqref{WV}.
We further require the conditions $H^{\text{tot},a}_0 = 0$ and $L_0^\text{tot}=0$ corresponding to \eqref{Jcond} and \eqref{Ccond}. Then the vertex operator can be restricted to the form 
\begin{align}
	V = e^{b j \cdot \varphi} e^{ i s \cdot Y} e^{ \tilde  b  (2 \rho - j  + s) \cdot  \tilde \varphi} \, ,
\end{align}
where $\rho$ is the Weyl vector defined in \eqref{Weyl}.

The effective action to evaluate correlation functions is 
\begin{align}
	\begin{aligned}
		S & = \frac{1}{ 4 \pi} \int d^2 w \left[ \partial \varphi \cdot \bar \partial \varphi + \partial Y \cdot \bar \partial Y 
		+ \partial \tilde \varphi \cdot  \bar \partial \tilde \varphi+ \frac{\sqrt{g} \mathcal{R} }{2} \sum_{a=1}^{n-1} (b \phi^a + \tilde b \tilde \phi^a)   \right]  \\
		& \quad +	\frac{ \lambda }{ 2 \pi} \int d^2 w \sum_{j=1}^{n-1} e^{ b (\varphi_{j} - \varphi_{j+1}) + i (Y_{j} - Y_{j+1}) }  \, .
	\end{aligned}
\end{align}
We rotate the fields as
\begin{align}
	 b \varphi_j + i Y_j = b ' \varphi _j ' \, , \quad - i  \varphi +  b Y = b ' Y ' \, , \quad b' = \sqrt{\frac{1 + n -k }{k - n}} \, .
\end{align}
We further define $\phi '_a = \varphi '_{a} - \varphi '_{a+1}$ and $\mathcal{Y} _a = Y'_{a} - Y '_{a +1}$.
Here we should notice that $\sum_{l=1}^n Y_l ' $ decouples as well.
The  action is now 
\begin{align}
\begin{aligned}
		S & = \frac{1}{ 2 \pi} \int d^2 w \left[ \frac{G^{(n)}_{ab}}{2} \left( \partial \phi '{}^a \bar \partial \phi '{}^b + \partial \mathcal{Y}^a  \bar \partial \mathcal{Y}^b  
		+ \partial \tilde \phi^a  \bar \partial \tilde \phi^b  \right)  \right] \\
		& \quad +	\frac{ \lambda}{ 2 \pi} \int d^2 w \left[  \frac{\sqrt{g} \mathcal{R} }{ 4 } \sum_{a=1}^{n-1} \left( Q_{\phi ' } \phi ' {}^a + Q_{\mathcal{Y} } \mathcal{Y}' {}^a + Q_{\tilde \phi  } \tilde \phi {}^a \right) + \sum_{a=1}^{n-1} e^{ b ' \phi ' _a}  \right]
\end{aligned}
\end{align}
with the background charges
\begin{align}
	Q_{\phi '} =  b' + 1/b'  \, , \quad Q_{Y '} = - i \tilde b \, , \quad Q_{\tilde \phi} = \tilde b \, .
\end{align}
The vertex operator \eqref{Vnew} is
\begin{align}
	V = e^{( (b' + 1/b') j - 1/b ' s) \cdot \varphi '} e^{ -  i  \tilde b (j- s) \cdot Y ' } e^{\tilde b (2 \rho - j + s) \cdot \tilde \varphi  } \, .
\end{align}
The contributions from $Y'$ and $\tilde \varphi ' $ cancel out up to reflection relations, which effectively project the field space spanned by $\varphi_j , Y_j$ to that spanned only by $\varphi_j '$.
We then arrive at the correlation function
\begin{align}
	\left \langle \prod_{\nu=1} ^N V_\nu (z_\nu) \right \rangle \, , \quad 	V_\nu (z_\nu)= e^{((b' + 1/b') j_\nu - 1/b ' s_\nu) \cdot \varphi ' (z_\nu)}  \label{todacorr}
\end{align}
with the action of $\mathfrak{sl}(n)$ Toda field theory as
\begin{align}
	S [\phi ' , \psi ] = \frac{1}{ 2 \pi} \int d^2 w \left[ \frac{G^{(n)}_{ab}}{2} \partial \phi ' {}^a  \bar \partial \phi ' {}^b  + \frac{\sqrt{g} \mathcal{R} }{4 }(b' +1/b') \sum_{a=1}^{n-1}\phi ' {}^a  
	+  \lambda \sum_{a=1}^{n-1} e^{ b ' \phi ' {}_a } \right] \, .  \label{todaaction}
\end{align}
In this way, we have shown that the $N$-point functions of primary operators in the BRST formulation of the coset can be reduced to the $N$-point functions as in \eqref{todacorr} with the action of $\mathfrak{sl}(n)$ Toda field theory \eqref{todaaction}.

\subsection{$\mathcal{N}=1$ super Liouville theory from coset model}
\label{sec:N=1}

We then examine the $N$-point functions of the coset
\begin{align}
	\frac{SL(2)_{k} \otimes SL(2)_{-2}}{SL(2)_{k - 2}} 
	\label{N1Vir}
\end{align}
and reduce them to those of $\mathcal{N}=1$ super Liouville theory.
The factor $SL(2)_{-1}$ in the coset \eqref{Vircoset} is now replaced by $SL(2)_{-2}$, which can be described by adjoint fermions $\psi^\pm , \psi^3$. The action for the coset is almost same as \eqref{BRSTaction} and given by
\begin{align}
	S = S_k^\text{WZNW} [\phi, \beta , \gamma ] + S _\psi [\psi] + S_{- k + 6}^\text{WZNW} [\tilde \phi , \tilde \beta , \tilde \gamma] + S_{bc} [b^a ,c_a ] \, .
\end{align}
The differences are the second action for the adjoint fermions, which may be expressed as
\begin{align}
	S  _{\psi}[\psi] = \frac{1}{2 \pi} \int d^2 w  \left[  \psi^+ \bar \partial \psi^- + \frac12  \psi^3 \bar \partial \psi^3 +  \bar \psi^+ \partial \bar \psi^- + \frac12  \bar \psi^3  \partial \bar \psi^3  \right] 
\end{align}
and the shift of level for the third action.
The conformal weights of fermions are $1/2$ and the fermions are bosonized by
\begin{align}
	\psi^\pm = e^{\pm i \sqrt{2} H^L} \, , \quad H^L (z) H^L (0) \sim-  \frac12 \ln z 
\end{align}
and similarly for $\bar \psi^\pm$ written by $H^R$. We further define $H = H^L + H^R$.
The $\mathfrak{sl}(2)$ current generators are given by
\begin{align}
	J^+_\psi = \sqrt{2} \psi^+ \psi^3  \, , \quad J^3_\psi = \psi^+ \psi^- \, , \quad J^-_\psi = \sqrt{2} \psi^- \psi^3 \, .
\end{align}
With this replacement,  the BRST charge is of  the same form as \eqref{BRSTcharge}.

Let us consider the case where all vertex operators are in the NSNS-sector.
Then the vertex operators can be put in the form 
\begin{align}
	V = \mathcal{P} (\gamma , \tilde \gamma) e^{2 b j \phi} e^{ i \sqrt{2}  s H } e^{2 \tilde b \tilde\jmath \tilde \phi } \label{ops}
\end{align}
as in \eqref{op} but now with $\tilde b = 1/\sqrt{- k + 4}$.
We further replace $\beta$ in $S_k^\text{WZNW}$ by
\begin{align}
	\beta (w)- \oint_w \frac{dz}{2 \pi i}  b^+ (z) Q (w) = \sqrt{2} \psi^3 e^{ i\sqrt{2}  H^L}  (w) - \tilde \beta (w) + \sqrt{2} ( b^+ (w) c_3 (w) + b^3(w) c_- (w)) \, .  
\end{align}
Then, we can use a non-trivial element of $Q_1$-cohomology as a vertex operator such as
\begin{align}
	V = e^{2 b j \phi} e^{ i \sqrt{2} s H} e^{2 \tilde b (1 - j + s) \tilde \phi } \label{Vnews}
\end{align}
as before. Here we have used the constraints as in \eqref{Virconst}.

The effective action to evaluate correlation functions is 
\begin{align}
	\begin{aligned}
	S & = \frac{1}{ 2 \pi} \int d^2 w \left[ \partial \phi \bar \partial \phi + \partial H \bar \partial H 
	+ \partial \tilde \phi \bar \partial \tilde \phi+ \frac{\sqrt{g} \mathcal{R} }{4} (b \phi + \tilde b \tilde \phi)  + \frac{1}{2} (\psi \bar \partial \psi + \bar \psi \partial \bar \psi)  \right]  \\
& \quad +	\frac{ \lambda }{  \pi} \int d^2 w\  \psi \bar \psi e^{2 b \phi + i \sqrt{2}  H }  \, ,
  \end{aligned}
\end{align}
where $\psi = \psi^3$ and $\bar \psi = \bar \psi^3$.
Rotating the fields as
\begin{align}
	2 b \phi + i \sqrt{2} H = b ' \phi ' \, , \quad - i \sqrt{2} \phi + 2 b H = b ' H ' \, , \quad \sqrt{2} \tilde \phi = \tilde \phi '
 \, , 
	\quad b' = \sqrt{\frac{4 -k}{k -2 }} \, ,
\end{align}
the  action is now 
\begin{align}
	\begin{aligned}
		S & = \frac{1}{ 4 \pi} \int d^2 w \left[ \partial \phi ' \bar \partial \phi ' + \partial H ' \bar \partial H '  
	+ \partial \tilde \phi ' \bar \partial \tilde \phi '  + \frac{\sqrt{g} \mathcal{R} }{4} \left( Q_{\phi '} \phi ' + Q_{H'} H' + Q_{\tilde \phi  '} \tilde \phi ' \right)  \right]  \\
		& \quad +	\frac{ 1}{ 4 \pi} \int d^2 w \left[ \psi \bar \partial \psi + \bar \psi \partial \psi + 4 \lambda \psi \bar \psi e^{ b ' \phi '} \right]  
	\end{aligned}
\end{align}
with the background charges
\begin{align}
	Q_{\phi '} =  b' + 1/b'  \, , \quad Q_{H '} = -  i \sqrt{2}\tilde b \, , \quad Q_{\tilde \phi '} =   \sqrt{2} \tilde b  \, .
\end{align}
The vertex operator \eqref{Vnews} becomes
\begin{align}
	V = e^{((b' + 1/b') j -1/ b ' s) \phi '} e^{ - i \sqrt{2}  \tilde b (j- s) H '} e^{\sqrt{2} \tilde b (1 - j + s) \tilde \phi  ' } \, .
\end{align}
For $N$-point function, we can see that the contributions from $H'$ and $\tilde \phi ' $ cancel out as before.
Therefore, we end up with the correlation function as
\begin{align}
	\left \langle \prod_{\nu=1} ^N V_\nu (z_\nu) \right \rangle \, , \quad 	V_\nu (z_\nu)= e^{((b' + 1/b') j_\nu -1/ b ' s_\nu) \phi ' (z_\nu)} \label{N1corr}
\end{align}
with the action of $\mathcal{N}=1$ super Liouville theory given by
\begin{align}
	S [\phi ' , \psi ] = \frac{1}{ 4  \pi} \int d^2 w \left[ \partial \phi ' \bar \partial \phi '  + \frac{\sqrt{g} \mathcal{R} }{4}(b' +1/b') \phi '  + \psi \bar \partial \psi + \bar \psi \partial \bar \psi 
	+ 4 \lambda \psi \bar \psi e^{ b ' \phi ' } \right] \, . \label{N1action}
\end{align}
In this way, we have reduced the $N$-point functions of primary operators in the BRST formulation of the coset \eqref{N1Vir} to the $N$-point functions \eqref{N1corr} with the action of $\mathcal{N}=1$ super Liouville theory \eqref{N1action}.

\section{Higher rank FZZ-duality}
\label{sec:HR}

In our previous paper \cite{Creutzig:2020cmn}, we examined higher rank FZZ-duality between the coset \eqref{hrcoset} and a theory with an $\mathfrak{sl}(n|n+1)$-structure. In that paper, however, we  only succeeded to derive correlator correspondences for the cases with $n=2,3$ due to our limited understanding of the methods of \cite{Gerasimov:1989mz,Kuwahara:1989xy}. In the previous  section, we established these methods for simple but important examples. In this section, we derive  correlator correspondences for higher rank FZZ-duality for all $n$ by applying the BRST-method to these examples.

\subsection{A first order formulation of coset model}

In order to specify which $(\beta , \gamma)$-systems cancel with each other, we have to choose a proper free field realization of the numerator algebra.
Namely, we construct a free field realization of affine $\mathfrak{sl}(n+1)$ such that the embedding of affine $\mathfrak{sl}(n)$ becomes simpler. We first review a free field realization used in subsection \ref{sec:hrg} for $\mathfrak{sl}(n)$ subalgebra with slightly different notation and then find out that for $\mathfrak{sl}(n+1)$ by extending it.

We introduced $n$ free bosons $\varphi_a$ and $n(n-1)/2$ pairs of $(\beta_{i,j},\gamma_{i,j})$-systems with $i  > j$.
The conformal weights of  $(\beta_{i,j},\gamma_{i,j})$ are $(1,0)$, respectively.
Among $n$ free bosons $\varphi_a$, one linear combination decouples though.
The non-trivial OPEs of these fields are
\begin{align}
	\varphi_a (z) \varphi_b (0) \sim - \delta_{a,b} \ln | z |^2 \, , \quad  \beta_{i,j} (z)  \gamma_{k,l} (0)\sim \frac{\delta_{i,k} \delta_{j,l}}{z} \, .
\end{align}
The currents $J^{\mathfrak{sl}(n)}_{i,j}$ with $i > j$ are constructed as 
\begin{align}
J^{\mathfrak{sl}(n)}_{i,j} = \beta_{i,j} - \sum_{l=1}^{j-1} \beta_{l , i}  \gamma_{l , j } \, .
 \end{align}
The Cartan subalgebra is generated by
\begin{align}
	H^{\mathfrak{sl}(n)}_a = \hat H^{\mathfrak{sl}(n)}_{a} -   \hat H^{\mathfrak{sl}(n)}_{a+1} \, , \quad 
\hat H^{\mathfrak{sl}(n)}_a =  - b_{(n)}^{-1} \partial \varphi_a - \sum_{l=1}^{a-1} \gamma_{a,l} \beta_{a,l} + \sum_{l=a+1
}^n \gamma_{l,a} \beta_{l, a} \, .
\end{align}
Here we set $b_{(n)} = 1/\sqrt{k-n}$.
The other generators $J^{\mathfrak{sl}(n)}_{i,j}$ $(i<j)$ are determined by requiring the correct OPEs with these currents.

We then look for a free field realization of affine $\mathfrak{sl}(n+1)$.
For this, we introduce a free boson $\varphi_{n+1}$ and $n$ additional pairs  $(\beta_{n+1,j},\gamma_{n+1,j})$ with $j=1,2,\ldots,n$ satisfying
\begin{align}
	\varphi_{n+1} (z) \varphi_{n+1} (0) \sim - \ln |z |^2\, , \quad \beta_{n+1,i}(z) \gamma_{n+1,j}(0) \sim \frac{\delta_{i,j}}{z} \, .
\end{align}
The conformal weights of   $(\beta_{n+1,j},\gamma_{n+1,j})$  are $(1,0)$, respectively.
We would like to obtain a special free field realization such that $J_{i,j} = J^{\mathfrak{sl}(n)}_{i,j}$ for $i > j$.
We find that 
\begin{align}
	J_{n+1,i} = \beta_{n+1,i}- \sum_{l = 1}^{i-1} \gamma_{i,l}  \beta_{n+1,l}
\end{align}
is consistent with the OPEs with $J_{i,j} = J^{\mathfrak{sl}(n)}_ {i,j}$. This also fixes the Cartan generators. Among them, we find
\begin{align}
	H_a = \hat H_{a} - \hat H_{a+1} \, , \quad  
\hat H_a = - b^{-1}_{(n+1)}\partial \varphi_a - \sum_{l=1}^{a-1} \gamma_{a,l} \beta_{a,l} + \sum_{l=a+1
}^{n+1} \gamma_{l,a} \beta_{l,a}    \, .
\end{align}
Here we set $b_{(n+1)}= 1/\sqrt{k-n-1}$.
We also have
\begin{align}
	Z = - \frac{1}{b_{(n+1)} (n+1)} \left( n \partial \varphi_{n+1} -  \sum_{a=1}^n \partial \varphi_a \right) -  \sum_{l=1
	}^{n} \gamma_{n+1,l} \beta_{n+1,l} \, .
\end{align}

The screening charges are also constructed such as to commute with these currents.
We find
\begin{align}
	Q_l = \int dz V_l(z)
\end{align}
with
\begin{align}
	V_l = \left( \beta_{l+1,l} - \sum_{j=l+2}^{n} \beta_{j,l} \gamma_{j,l+1} - \beta_{n+1,l} \gamma_{n+1,l+1} \right) e^{b_{(n+1)} (\varphi_l - \varphi_{l+1})}  
\end{align}
for $l=1,2,\ldots,n-1$ and
\begin{align}
	V_{n} = \beta_{n+1,n} e^{ b _ {(n+1)}(  \varphi_n - \varphi_{n+1} ) } \, .
\end{align}

We then move to find a first order formulation of the coset  \eqref{hrcoset} by applying the method developed in the previous section.
We have observed that this effectively reduces the method proposed by \cite{Gerasimov:1989mz,Kuwahara:1989xy} expect for the interaction terms.
Namely, we consider the field space orthogonal to the denominator factors $SL(n) \times U(1)$.
For this, we first neglect $(\beta_{i,j},\gamma_{i,j})$ without $\beta_{n+1,j} \, (\equiv \beta_j)$ and 
$\gamma_{n+1,j}\, (\equiv \gamma_j)$. We then introduce free bosons
\begin{align}
	\hat H_a = b^{-1}_{(n)}\partial \hat \varphi_a + \sum_{l=1}^{a-1} \gamma_{a,l} \beta_{a,l} - \sum_{l=a+1
	}^{n} \gamma_{l,a} \beta_{l,a}   \, , \quad
Z =  \sqrt{ \frac{k n}{n+1}} \partial \hat \varphi_{n+1}
\end{align}
with $a=1,2,\ldots, n$, and consider the orthogonal space to $\hat \phi_a = \hat \varphi_{a} - \hat \varphi_{a+1}$ $(a=1,2,\ldots , n-1)$ and $\hat \varphi_{n+1}$ as well.
Instead of doing so, we introduce $\chi_j$ for $j=1,2,\ldots , n -1$ and $\eta$ with the opposite sign in front of the kinetic terms.
The corresponding action is
\begin{align}
	\begin{aligned}
	 S &= \frac{1}{2 \pi} \int d^2 w \left[ \frac{G^{(n+1)}_{ab}}{2} \partial \phi^a \bar \partial \phi^b 
	 	- \frac{G^{(n)}_{ij}}{2} \partial \chi^i \bar \partial \chi^j - \frac12 \partial \eta \bar \partial \eta  - \sum_{j=1}^n \left(\beta_j \bar \partial \gamma _j+ \bar \beta_j  \partial \bar \gamma_j\right) \right]\\
	 	&  \quad + \frac{1}{2 \pi} \int d^2 w \left[ \frac{\sqrt{g} \mathcal{R}}{4} \left(b_{(n+1)} \sum_{a=1}^n \phi^a - b_{(n)} \sum_{j=1}^{n-1} \chi^j \right) + \lambda \sum_{l=1}^{n} V_l \right] \label{cosetaction}
\end{aligned}
\end{align}
with
\begin{align}
V_l = |\beta_{l} \gamma_{l+1} |^2 e^{b_{(n+1) } \phi_l} \quad (l =1,2,\ldots, n-1) \, , \quad 	V_n = |\beta_n|^2 e^{b_{(n+1)} \phi_n} \, .
\end{align}
Here we set $\phi_a = \varphi_{a } - \varphi_{a+1}$.
We consider the vertex operator of the form
\begin{align}
	\Psi (z) = \left[ \prod_{i=1}^n \gamma_i^{\alpha_i} \bar \gamma_i ^{\bar \alpha_i} \right] 
	e^{b_{(n+1)} \sum_{a=1}^{n} j_a \phi_a} e^{b_{(n)} \sum_{j=1}^{n-1} l_j \chi_j} e^{\sqrt{\frac{n (n+1)}{k}} (m \eta_L + \bar m \eta_R )} \, , \label{cosetvo}
\end{align}
where
\begin{align}
	\alpha_i = - j_{i} + j_{i-1} - l_{i} + l_{i-1} - m \, .
\end{align}
Here we set $j_{0} = l_{0} = l_n = 0$.
We define $\bar \alpha_i$ by replacing $m$ with $\bar m$.

\subsection{Application of reduction method}

Now that we have a first order formulation of the coset model \eqref{hrcoset}, we could apply the analysis in \cite{Hikida:2008pe,Creutzig:2020cmn}.
We consider the correlation function of the form
\begin{align}
\left \langle	\prod_{\nu=1}^N \Psi_\nu (z_\nu) \right \rangle \, .
\end{align}
We use the coset action \eqref{cosetaction} and the vertex operator of the form \eqref{cosetvo}.
It would be convenient to rewrite the vertex operators as
\begin{align}
	\Psi_\nu (z_\nu) = \Phi_\nu (z_\nu) V^{\chi;\eta}_\nu (z_\nu) \, , \quad 
	 V^{\chi;\eta}_\nu (z_\nu) = e^{b_{(n)} \sum_{j=1}^{n-1} l_j ^\nu\chi_j} e^{\sqrt{\frac{n (n+1)}{k}} (m ^\nu\eta_L + \bar m^\nu \eta_R )} \, ,
\end{align}
where $\Phi_\nu$ is defined as
\begin{align}
&	\Phi_\nu (z_\nu) = \int  \left[ \prod_{a=1}^n  \frac{d ^2 \mu_a^\nu}{|\mu_a^\nu|^2}  (\mu_a^\nu)^{ - j_{a} ^\nu + j_{a-1}^\nu - \alpha_a^\nu} (\bar \mu_a ^\nu ) ^{ - j_{a}^\nu + j_{a-1} ^\nu- \bar \alpha_a^\nu} \right]
	V_\nu (z_\nu) \, ,  \label{Phi}\\
&	V_\nu (z_\nu)  = \left[ \prod_{a=1}^n |\mu_a^\nu|^{ 2 ( j_{a} ^\nu - j_{a-1} ^\nu )}  \right]
	  e^{\sum_{a=1}^n (\mu^\nu_a \gamma_a - \bar \mu^\nu_a \bar \gamma_a )} e^{\sum_{a=1}^n j_a ^\nu \phi_a}
\end{align}
with $j_0 ^\nu =0$.
As closely explained in \cite{Hikida:2008pe,Creutzig:2020cmn}, we can introduce the identity operator in the coset model. In the current first order expression, it is given by
\begin{align}
\mathbbm{1} = v^{\{S_j\}} (\xi) e^{\sum_{j=1}^{n-1}(S_{j+1} - S_{j} ) \chi^j /b_{(n)}}e^{- \sqrt{\frac{k}{n(n+1)}} ( \sum_{j=1}^n S_j \eta)} \, , 
\end{align}
where $v^{\{S_j\}} (\xi)$ restricts the domain of integration over $\beta_j$ to have a zero of order $S_j$ and extra insertion $e^{\sum_{a=1}^n(S_{a} - S_{a+1} ) \phi^a /b_{(n+1)}}$ with $S_{n+1}=0 $ at $w =\xi$.
Moreover, we have set $\eta = \eta_L + \eta_R$.
If the interaction terms in  \eqref{cosetaction} do not include $\gamma_a$, then we can integrate it out and $\beta_a$ can be replaced by a function defined by
\begin{align}
\sum_{\mu=1}^N  \frac{\mu_a^\nu}{w - z_\nu} 
 = u_a \frac{(w - \xi)^{S_a}\prod_{p=1}^{N-2 -S_a} (w - y_p^a)}{\prod_{\nu=1}^N (w - z_\nu)} = u_a	\mathcal{B}_a \label{Ba}
\end{align}
subject to constraints
\begin{align}
\sum_{\nu=1}^n \frac{\mu^\nu_a}{(w - \xi)^n} = 0
\end{align}
 for $n=0,1,\ldots, S_a$.
Since it is not the case in general, we need some tricks as in \cite{Creutzig:2020cmn}.

From the interaction terms of the action \eqref{cosetaction}, we can see that $\gamma_1$ does not appear.
Thus we can integrate $\gamma_1$ out and $\beta_1$ can be replaced by a function $u_1 \mathcal{B}_1$ defined in  \eqref{Ba}. We shift the fields as
\begin{align}
	\begin{aligned}
&\phi_1 + \frac{1}{b_{(n+1)}}| u_1 \mathcal{B}_1|^2 \to \phi_1 \, , \\
&\chi_{1} + \frac{1}{b_{(n)}}| u_1 \mathcal{B}_1|^2 \to \chi_{1} \, ,  \\
& \eta_L + \sqrt{\frac{k}{n (n+1)}}| u_1 \mathcal{B}_1|^2 \to \eta_L
 \end{aligned}
\end{align}
and similarly for $\eta_R$ as in the previous works.
Essentially there are two contributions coming from kinetic terms. The first one is the change of vertex operator with the removal of $\mu_1^\nu$ and the shift of parameters as
\begin{align}
	V_\nu (z_\nu)  \to  V_\nu (z_\nu) = \left[ \prod_{a=2}^{n} |\mu^\nu_a|^{ 2 ( j_{a} ^\nu - j_{a-1} ^\nu ) }  \right]
e^{\sum_{a=2}^{n} (\mu^\nu_a \gamma_a - \bar \mu^\nu_a \bar \gamma_a )} e^{b_{(n+1)}\sum_{a=1}^n j_a ^\nu \phi_a + \phi^1 /b_{(n+1)}}
\end{align}
and
\begin{align}
V^{\chi;\eta}_\nu (z_\nu) \to 	 V^{\chi;\eta}_\nu (z_\nu) = e^{b_{(n)} \sum_{j=1}^{n-1} l_j ^\nu \chi_j - \chi^{1}/b_{(n)}} e^{\sqrt{\frac{n (n+1)}{k}} ( ( m ^\nu -  \frac{k}{n (n+1)} )  \eta_L + ( \bar m ^\nu -  \frac{k}{n (n+1)} ) \eta_R )} \, .
\end{align}
The other is the extra insertion of vertex operators 
\begin{align}
	V_b (y_p^1) = e^{- \phi^1 / b_{(n+1)} + \chi^1 / b_{(n)} + \sqrt{\frac{k}{n (n+1)}}  \eta} 
\end{align}
 for $p=1,2,\ldots , N-2 - S_1$. 
We regard this term as an interaction term as in \cite{Hikida:2008pe,Creutzig:2020cmn}.
This is possible since the integration over $\mu_1^\nu$ in \eqref{Phi} becomes that over $y_p^1$ via the change of variables \eqref{Ba}.

We can integrate $\gamma_1$ out, but we cannot do so $\gamma_a$ with $a \neq 1$ at least naively.
Here we take a route to treat an interaction term perturbatively as suggested in \cite{Creutzig:2020ffn}.
Let us focus on the following two interaction terms;
\begin{align}
	V_{2} = |\beta_{2} \gamma_{3} |^2 e^{b_{(n+1)} \phi_{2}} \, , \quad 
	V_{1} = |\gamma_{2}|^2 e^{b_{(n+1)} \phi_{1}} \, .
\end{align}
As explained in appendix \ref{sec:BP}, we can change the vertex operator with $\gamma_{2}$ to that without $\gamma_{2}$ by field redefinitions. Then, we perform a self-duality of Liouville field theory with the interaction term. After that, we come back to the original fields. The interaction term $V_{1} $ now becomes%
\footnote{Here the formula is up to overall factor. The same is true for other cases.}
\begin{align}
	\begin{aligned}
		V_{1} &= |\gamma_{2}|^{2 (k-n-1)} e^{ \phi_{1} /b_{(n+1)}} \\
		 &= 
		\int \frac{ d ^2 \mu_{2}}{ |\mu_{2}|^2}  |\mu_{2}|^{2(- k+ n+1)} 
		e^{\mu_{2} \gamma_{2} - \bar \mu_{2} \bar \gamma_{2}  }e^{ \phi_{1} /b_{(n+1)}}  \, .
		\end{aligned}
\end{align}
Treating this term perturbatively, we can integrate $\gamma_{2}$ out and $\beta_{2}$ is replaced by a function $u_{2} \mathcal{B}_{2}$ defined in \eqref{Ba}.
In order to remove the function (or $\mu_{2}^\nu$), we shift the variables as
\begin{align}
	\begin{aligned}
		&\phi_{1} - \frac{1}{b_{(n+1)}}| u_{2} \mathcal{B}_{2}|^2 \to \phi_{1} \, , \quad
		\phi_{2} + \frac{1}{b_{(n+1)}}| u_{2} \mathcal{B}_{2}|^2 \to \phi_{2} \, , \\
		&\chi_{1} - \frac{1}{b_{(n)}}| u_{2} \mathcal{B}_{2}|^2 \to \chi_{1} \, ,  \quad
		\chi_{2} + \frac{1}{b_{(n)}}| u_{2} \mathcal{B}_{2}|^2 \to \chi_{2} \, ,  \\
		& \eta_L + \sqrt{\frac{k}{n (n+1)}} u_2 \mathcal{B}_2  \to \eta_L
	\end{aligned}
\end{align}
and similarly for $\eta_R$. The term $V_{1}$ becomes 
\begin{align}
	\begin{aligned}
	V_{1} \to V_{1} &= e^{ ( \phi_{1}  - \phi^{1} + \phi^{2}) / b_{(n+1)} + ( \chi^{1} - \chi^{2}) / b_{(n)} - \sqrt{\frac{k}{n (n+1)}}  \eta}  \\
&	= e^{ \phi^{1}  / b_{(n+1)} + ( \chi^{1} - \chi^{2}) / b_{(n)} - \sqrt{\frac{k}{n (n+1)}}  \eta} \, , 
    \end{aligned}
\end{align}
which can be put back as one of interaction terms.
The change of vertex operator  with the removal of $\mu_{2}^\nu$ and the shift of parameters is
\begin{align}
	V_\nu (z_\nu)  \to  V_\nu (z_\nu) = \left[ \prod_{a=3}^{n } |\mu_a^\nu |^{ 2 ( j_{a}^\nu - j_{a-1} ^\nu ) }  \right]
	e^{\sum_{a=3}^{n} (\mu^\nu_a \gamma_a - \bar \mu^\nu_a \bar \gamma_a )} e^{b_{(n+1)}\sum_{a=1}^n  j_a ^\nu \phi_a + \phi^{2} /b_{(n+1)}}
\end{align}
and
\begin{align}
	V^{\chi;\eta}_\nu (z_\nu) \to 	 V^{\chi;\eta}_\nu (z_\nu) = e^{b_{(n)} \sum_{j=1}^{n-1} l_j ^\nu \chi_j - \chi^{2}/b_{(n)}} e^{\sqrt{\frac{n (n+1)}{k}} ( ( m ^\nu -  \frac{2 k}{n (n+1)} )  \eta_L + ( \bar m ^\nu -  \frac{2 k}{n (n+1)} ) \eta_R )} \, .
\end{align}
The other is the extra insertions of vertex operators 
\begin{align}
	V_b (y_p^2) = e^{ (\phi^1 - \phi^{2}) /b_{(n+1)} - ( \chi^{1} - \chi^{2} ) / b_{(n)} + \sqrt{\frac{k}{n (n+1)}}  \eta} 
\end{align}
for $p=1,2,\ldots , N-2 - S_2$, which are regarded as interaction terms.

In a similar manner, we can integrate $\gamma_a$ for all $a$.
The vertex operator is now 
\begin{align}
&	\Psi_\nu (z_\nu)  \to 	\Psi_\nu (z_\nu)   =   e^{b_{(n+1)}\sum_{a=1}^n j^\nu _a \phi_a + \phi^{n} /b_{(n+1)} + b_{(n)} \sum_{j=1}^{n-1} l_j ^\nu \chi_j + \sqrt{\frac{n (n+1)}{k}} ( ( m^\nu - \frac{k}{n+1} )  \eta_L + ( \bar m ^\nu - \frac{k}{n+1} ) \eta_R )} \, .
\end{align}
The interaction terms are 
\begin{align}
	\begin{aligned}
	V_{l} &= e^{( \phi_l - \phi^{l} + \phi^{l+1} ) / b_{(n+1)} + ( \chi^{l} - \chi^{l+1}) / b_{(n)} - \sqrt{\frac{k}{n(n+1)} } \eta} \\
	&= e^{- ( \phi^{l-1} -  \phi^{l} ) / b_{(n+1)} + ( \chi^{l} - \chi^{l+1}) / b_{(n)} - \sqrt{\frac{k}{n(n+1)} } \eta}  
    \end{aligned}
\end{align}
for $l=2,3,\ldots , n-2$ and
\begin{align}
	V'_{l} = e^{( \phi^{l-1} - \phi^{l} ) / b_{(n+1)} - (  \chi^{l-1} - \chi^{l}) / b_{(n)} + \sqrt{\frac{k}{n(n+1)} }\eta} 
\end{align}
for $l=2,3,\ldots , n-1$. Moreover, we have
\begin{align}
	\begin{aligned}
	   &V_{1} = e^{ \phi^1  / b_{(n+1)} + ( \chi^{1} - \chi^2 )  / b_{(n)} - \sqrt{\frac{k}{n(n+1)} } \eta}  \, , \\
		&V_{n-1} = e^{ - (\phi^{n-2} - \phi^{n-1})   / b_{(n+1)} + \chi^{n-1}   / b_{(n)} - \sqrt{\frac{k}{n(n+1)} } \eta}  \, , \\
		&V_n = e^{\phi_n/b_{(n+1)}} 
    \end{aligned}
\end{align} 
and
\begin{align}
	\begin{aligned}
&	V'_{1} = e^{ - \phi^1 / b_{(n+1)} +  \chi^{1}  / b_{(n)} + \sqrt{\frac{k}{n(n+1)}} \eta} \, , \\
&	V'_n = e^{( \phi^{n-1}- \phi^{n} ) /b_{(n+1)} - \chi^{n-1} / b_{(n)} + \sqrt{\frac{k}{n (n+1)}}  \eta}  \, .
  \end{aligned}
\end{align}
For $V_n$, we have performed a self-duality of Liouville field theory.
The kinetic terms are similar to those of \eqref{cosetaction}.
Only differences are no $(\beta_i , \gamma_i)$ now and the shifts of background charges for $\phi^n$ and $\eta$ as
\begin{align}
	Q_{\phi^n} = b_{(n+1)} + b_{(n+1)}^{-1} \, , \quad Q_\eta = \sqrt{\frac{k n}{n+1}} \, .
\end{align}

\subsection{Structure of the dual theory}

As explained in  \cite{Creutzig:2020cmn} (see also \cite{BFM,Litvinov:2016mgi,Prochazka:2018tlo}), 
the symmetry algebra of the dual theory should be given by $Y_{n,0,n+1}[\psi^{-1}]$-algebra with $\psi = - k + n + 1$ after decoupling a $\mathfrak{u}(1)$ subalgebra. In order to express it, we introduce $\phi^{(1)}_j$  with $j=1,2,\cdots ,n$ and $\phi^{(3)}_j$ with $j=1,2,\cdots , n+1$. The normalization is
\begin{align}
	\phi^{(1)}_j (z) \phi^{(1)}_l (0) \sim - \frac{1}{h_2 h_3} \delta_{j,l} \ln z \, , \quad
		\phi^{(3)}_j (z) \phi^{(3)}_l (0) \sim - \frac{1}{h_1 h_2} \delta_{j,l} \ln z 
\end{align} 
with
\begin{align}
	h_1 = i \sqrt{k - n - 1} \, , \quad h_2 = \frac{i}{\sqrt{k - n -1}} \, , \quad h_3 = - i \frac{k-n}{\sqrt{j-n-1}} \, .
\end{align}
We may consider the free field realization corresponding to the ordering 
\begin{align}
	\phi_1^{(1)} \phi^{(3)}_1 \phi^{(1)}_2 \cdots \phi^{(1)}_{n} \phi^{(3)}_{n} \phi^{(3)}_{n+1} \, .
\end{align}
The screening operators are
\begin{align}
	\begin{aligned}
&	V '_ l = e^{- h_3 \phi^{(1)}_l + h_1 \phi^{(3)}_l}  \quad (l=1,2,\ldots , n) \, , \\
&	V_l = e^{- h_1 h^{(3)}_l + h_3 \phi^{(1)}_{l+1} }\quad (l=1,2,\ldots , n-1) \, , \quad 
	V_n = e^{- h_1 (h^{(3)}_n - \phi^{(3)}_{n+1} ) } \, .
	\end{aligned}
\end{align}
We can check that they reproduce the interaction terms obtained in the previous subsection using the coordinate transformation  in  \cite{Creutzig:2020cmn}. The Gram matrix has a correspondence to the Dynkin diagram of $\mathfrak{sl}(n+1|n)$, whose simple root system consists of $2n -1$ fermionic roots and one bosonic root.

It might be useful to map to the free field realization given in \cite{Creutzig:2020cmn}.
For this we replace $\phi^{(\kappa)}_j$ by $\phi ' {}^{(\kappa)}_j$.
The ordering is denoted as
\begin{align}
 \phi ' {}^{(3)}_{1} 	\phi ' {}_1^{(1)} \phi ' {}^{(3)}_2 \cdots  \phi ' {}^{(3)}_{n}  \phi ' {}^{(1)}_{n} \phi ' {}^{(3)}_{n+1} \, .
\end{align}
The screening operators are
\begin{align}
			V '_ l = e^{- h_1 \phi ' {}^{(3)}_l + h_3 \phi ' {}^{(1)}_l}  \, , \quad
		V_l = e^{- h_3 h ' {}^{(1)}_l + h_1 \phi ' {}^{(3)}_{l+1} }
\end{align}
for $l=1,2,\ldots , n$. Its Gram matrix 
has a correspondence to the Dynkin diagram of $\mathfrak{sl}(n+1|n)$, whose simple root system consists of $2n$ fermionic roots. We can check that this map can be done by applying the reflection relation of \cite{Hikida:2008pe} to the interaction terms. Explicitly, we need to use the reflection to $V_n$ w.r.t. $V'_n$ and $V_l$ w.r.t. $V'_{l}$ and $V'_{l+1}$ and act a rotation of fields.

The current vertex operator may be put into the form%
\footnote{Using the decoupled $U(1)$, we have set $a^{(3)}_{n+1} = 0$.}
\begin{align}
V =	e^{\sum_{j=1}^n a^{(1)}_j \phi^{(1)}_j + \sum_{l=1}^{n+1} a^{(3)}_l \phi^{(3)}_l} \, .
\end{align}
Applying the reflections by $V'_l$ for all $l$ and acting the same rotation of fields, the vertex operator can be mapped to 
\begin{align}
	V =	e^{\sum_{j=1}^n a '{}^{(1)}_j \phi ' {}^{(1)}_j + \sum_{l=1}^{n+1} a ' {}^{(3)}_l \phi ' {}^{(3)}_l} \, .
\end{align}
The momenta $a^{(\kappa)}_j$ and $a ' {}^{(\kappa)}_j$ are related as
\begin{align}
	a ' {}^{(1)}_j = a^{(1)}_j + i \frac{k-n}{\sqrt{k-n-1}} \, ,\quad  a ' {}^{(3)}_ j = a^{(3)}_j + i \sqrt{k- n-1} \, .
\end{align}
Going back to the original coordinate system, we find
\begin{align}
	&	\Psi_\nu (z_\nu)   =   e^{b_{(n+1)}\sum_{a=1}^n j_a ^\nu  \phi_a + b_{(n)} \sum_{j=1}^{n-1} l_j ^\nu \chi_j  + \sqrt{\frac{n (n+1)}{k}} ( m ^\nu  \eta_L +  \bar m ^\nu  \eta_R )}
\end{align}
as one may have expected.

\section{Fermionic higher rank FZZ-duality}
\label{sec:sHR}

In this section, we derive correlator correspondences between the $CP_n$ Kazama-Suzuki coset \eqref{KScoset}
and $\mathfrak{sl}(n|n+1)$ Toda field theory as conjectured in \cite{Ito:1990ac,Ito:1991wb}.
For small $n$, it was actually possible to derive correlator correspondences if the trick used around \eqref{kin} was recognized. Anyway, it was not possible to analyze the cases with generic $n$ before elaborating the method of \cite{Gerasimov:1989mz,Kuwahara:1989xy}. In particular, we have learned how to incorporate fermions in the interaction terms as was done in subsection \ref{sec:N=1}.

\subsection{A first order formulation of super coset model}

We describe the coset \eqref{KScoset} in the BRST formulation by applying the method developed in section \ref{sec:toda}.
The factor $SO(2n)_1$ can be described by Dirac fermions  $\psi^\pm_j$ $(j=1,2,\ldots , n)$ with conformal weight $1/2$.
The generators of $\mathfrak{sl}(n)_{-1}$ can be obtained by these fermions as
\begin{align}
	J^\psi_{i,j} = \psi_j^+ \psi_i^- \quad (i > j) \, , \quad \hat H^\psi_a = - \psi_a^+ \psi_a^- \, , 
\end{align}
and similarly for $J^\psi_{i,j}$ with $i  < j$. We frequently use its bosonized formulation as
\begin{align}
	\psi^\pm_j = e^{\pm i Y^L_j} \, , \quad Y^L_i (z) Y^L_j (0) \sim-  \delta_{i,j} \ln z \, .
\end{align}
We also introduce $Y^R_j$ in a similar manner.
For the super coset, we neglect $(\beta_{i,j},\gamma_{i,j})$ except for 
$\beta_{n+1,j} \equiv \beta_j$ and $\gamma_{n+1,j} \equiv \gamma_j$ as in the case of bosonic coset. However, in the current case, we need to introduce slightly different bosons as 
\begin{align}
 &\hat H_a +  \hat H^\psi_a = b^{-1} \partial \hat \varphi_a 
 + \sum_{l=1}^{a-1} \gamma_{a,l} \beta_{a,l} - \sum_{l=a+1}^n \gamma_{l,a} \beta_{l,a} \, , \\
& Z + \sum_{l=1}^n \psi_l^+ \psi_l^- = b^{-1} \sqrt{\frac{n}{n+1}} \partial \hat \varphi_{n+1} \, .  
\end{align}
Here and in the following, we set $b = b_{(n+1)}$. We then consider the orthogonal space to $\hat \phi_a = \hat \varphi_a -\hat  \varphi_{a+1}$ $(a=1,2,\ldots, n-1)$ and $\hat \varphi_{n+1}$. As before, we instead introduce $\chi_j$ for $j=1,2,\ldots ,n-1$ an $\eta$ with the opposite sign in the kinetic terms.

The corresponding action is now
\begin{align}
		S &= \frac{1}{2 \pi} \int d^2 w \left[ \frac{G^{(n+1)}_{ab}}{2} \partial \phi^a \bar \partial \phi^b 
		- \frac{G^{(n)}_{ij}}{2} \partial \chi^i \bar \partial \chi^j - \frac12 \partial \eta \bar \partial \eta  + \frac{ b \sqrt{g} \mathcal{R}}{4} \left(\sum_{a=1}^n \phi^a -  \sum_{j=1}^{n-1} \chi^j \right) \right] \nonumber \\
		& \quad +\frac{1}{2 \pi} \int d^2 w \left[  \sum_{j=1}^n \left(- \beta_j \bar \partial \gamma _j - \bar \beta_j  \partial \bar \gamma_j  + \psi^+_j \bar \partial \psi_j^- + \bar \psi^+_j  \partial \bar \psi^-_j\right) + \lambda \sum_{l=1}^{n} V_l \right]  \label{scosetaction}
\end{align}
with
\begin{align}
	V_l = |\psi^+_l \psi^-_{l+1} + \beta_{l} \gamma_{l+1} |^2 e^{b_{(n+1) } \phi_l} \quad (l =1,2,\ldots, n-1) \, , \quad 	V_n = |\beta_n |^2 e^{b_{n+1} \phi_n} \, .
\end{align}
As in subsection \ref{sec:hrg}, $\beta_{l+1,l}$ is replaced by the expression in \eqref{betaij}, which yields the terms involving fermions $\psi^+_l \psi^-_{l+1}$, and the terms including $\gamma_{i,j}$ except for $i = n+1$ are neglected.
Here vertex operators are assumed to be independent of BRST ghosts as well.
More precisely speaking, we consider the vertex operator of the form%
\footnote{The form of $\Gamma_i$ may be explained  as an eigenfunction of Laplace operator as in \cite{Hikida:2007sz}.}
\begin{align}
	\Psi (z) = \left[ \prod_{i=1}^n \Gamma_i^{\alpha_i} \bar \Gamma_i ^{\bar \alpha_i} \right] 
	e^{b ( \sum_{a=1}^{n} j_a \phi_a +  \sum_{j=1}^{n-1} l_j \chi_j + \sqrt{n (n+1)} (m \eta_L + \bar m \eta_R ) ) } \, , \label{scosetvo}
\end{align}
where 
\begin{align}
	\Gamma_1 = \gamma_1 \, , \quad 
	\Gamma_i = \gamma_i + \psi^+_{i-1} \psi^-_i  \quad (i=2,3,\ldots ,n) \, .
\end{align}
Moreover, we set
\begin{align}
	\alpha_i = - j_{i} + j_{i-1} - l_{i} + l_{i-1} - m 
\end{align}
and similarly for $\bar \alpha_i$ as before.
Here we have set $j_{0} = l_{0} = l_n = 0$.

\subsection{Application of reduction method}

Now we have a first order formulation of the Kazama-Suzuki coset \eqref{KScoset}, we can apply the reduction methods developed in \cite{Hikida:2008pe,Creutzig:2010bt,Creutzig:2020cmn}.
We consider the correlation function of the form
\begin{align}
	\left \langle	\prod_{\nu=1}^N \Psi_\nu (z_\nu) \right \rangle \, .
\end{align}
The effective action is given by \eqref{scosetaction} and
the vertex operators are rewritten as
\begin{align}
	\Psi_\nu (z_\nu) = \Phi_\nu (z_\nu) V^{\chi;\eta}_\nu (z_\nu) \, , \quad 
	V^{\chi;\eta}_\nu (z_\nu) = e^{b ( \sum_{j=1}^{n-1} l^\nu_j \chi_j + \sqrt{n (n+1)} (m^\nu \eta_L + \bar m^\nu \eta_R ))} \, ,
\end{align}
where $\Phi_\nu$ is defined as
\begin{align}
	&	\Phi_\nu (z_\nu) = \int  \left[ \prod_{a=1}^n \frac{ d ^2 \mu_a^\nu}{ |\mu_a^\nu|^2} \mu_a^{ - j_{a}^\nu + j_{a-1}^\nu  - \alpha_a^\nu } \bar \mu_a ^{ - j_{a} ^\nu + j_{a-1} ^\nu - \bar \alpha_a ^\nu } \right]
	V_\nu (z_\nu) \, , \\
	&	V_\nu (z_\nu)  = \left[ \prod_{a=1}^n |\mu_a^\nu |^{ 2 ( j_{a} ^\nu- j_{a-1}^\nu )}  \right]
	e^{\sum_{a=1}^n (\mu^\nu_a \Gamma_a - \bar \mu^\nu_a \bar \Gamma_a )} e^{\sum_{a=1}^n j_a ^\nu \phi_a}
\end{align}
with $j^\nu_0 =0$. We could insert an identity operator as in the bosonic case. In the current case, it can be expressed as
\begin{align}
\mathbbm{1} = v^{\{S_j\}} (\xi) e^{\sum_{j=1}^{n-1}(S_{j+1} - S_{j} ) \chi^j /b} |e^{i \sum_{j=1}^nY^L_j}|^2e^{-  \frac{1}{\sqrt{n(n+1)}} ( \sum_{j=1}^n S_j (\eta_L + \eta_R))/b} \, .
\end{align}
As before the spectral flow operator $v^{\{S_j\}} (\xi)$ restricts the domain of integration over $\beta_j$ to have a zero of order $S_j$ and extra insertion $e^{\sum_{a=1}^n(S_{a} - S_{a+1} ) \phi^a /b}$ with $S_{n+1}=0 $ at $w =\xi$.

Since $\Gamma_1 = \gamma_1$, integration over $\gamma_1,\beta_1$ can be done as in the bosonic case.
We shift the fields
\begin{align}
	\begin{aligned}
		&\phi_1 + \frac{1}{b}| u_1 \mathcal{B}_1|^2 \to \phi_1 \, , \quad
		\chi_{1} + \frac{1}{b}| u_1 \mathcal{B}_1|^2 \to \chi_{1} \, ,  \\
		& \eta_L + \frac{1}{b}\frac{1}{\sqrt{n (n+1)}}| u_1 \mathcal{B}_1|^2 \to \eta_L \, , \quad
			Y_1^L + i  \ln u_1 \mathcal{B}_1 \to  Y_1^L 
	\end{aligned}
\end{align}
and similarly for $\eta_R,Y_1^R$.
Essentially there are two contributions. The vertex operators are changed as
\begin{align}
	V_\nu (z_\nu)  \to  V_\nu (z_\nu) = |e^{ i Y_1^L} |^2 \left[ \prod_{a=2}^{n} |\mu_a|^{ 2 ( j_{a} - j_{a-1} ) }  \right]
	e^{\sum_{a=2}^{n} (\mu_\nu^a \Gamma_a - \bar \mu_\nu^a \bar \Gamma_a )} e^{b\sum_{a=1}^n j^\nu_a \phi_a + \phi^1 /b}
\end{align}
and
\begin{align}
	V^{\chi;\eta}_\nu (z_\nu) \to 	 V^{\chi;\eta}_\nu (z_\nu) = e^{b \sum_{j=1}^{n-1} l_j ^\nu \chi_j - \chi^{1}/b} e^{b \sqrt{n (n+1)} ((  m ^\nu -  \frac{1}{n (n+1) b^2 } )  \eta_L + ( \bar m ^\nu -  \frac{1}{n (n+1) b^2} ) \eta_R )} \, .
\end{align}
The other is the extra insertions of vertex operators 
\begin{align}
	V_b (y_p^1) = e^{ ( - \phi^1  + \chi^1  + \frac{1}{\sqrt{n (n+1)}}  \eta ) / b } | e^{- i Y_1^L}|^2 
\end{align}
 for $p=1,2,\ldots , N-2 - S_1$. 

Now the interaction term $V_1$ becomes
\begin{align}
	V_1 = | \psi_1^+ \psi_2^- + \gamma_2|^2 e^{b \phi_1} \, .
\end{align}
We may change the variable as
\begin{align}
 \Gamma_2 =  \gamma_2 + \psi_1^+ \psi_2^- \to \gamma_2 \, .
\end{align}
This changes $V_1$ as 
\begin{align}
	V_1 = |  \gamma_2|^2 e^{b \phi_1} \, ,
\end{align}
but the kinetic term becomes
\begin{align}
	- \beta_2 \bar \partial \gamma_2 \to - \beta_2 \bar \partial (\gamma_2 - \psi_1^+ \psi_2^-  )
	=  - \beta_2 \bar \partial \gamma_2 - ( \bar \partial \beta_2 ) \psi_1^+ \psi_2^- \, .  \label{kin}
\end{align}
The final equality is up to total derivative.
Now we can proceed as in the bosonic case.
We shift the variables as
\begin{align}
	\begin{aligned}
		&\phi_{1} - \frac{1}{b}| u_{2} \mathcal{B}_{2}|^2 \to \phi_{1} \, , \quad
		\phi_{2} + \frac{1}{b}| u_{2} \mathcal{B}_{2}|^2 \to \phi_{2} \, , \quad
		\chi_{1} - \frac{1}{b}| u_{2} \mathcal{B}_{2}|^2 \to \chi_{1} \, ,  \\
	&	\chi_{2} + \frac{1}{b}| u_{2} \mathcal{B}_{2}|^2 \to \chi_{2} \, ,  \quad
		 \eta_L + \frac{1}{b}\frac{1}{\sqrt{n (n+1)}} u_2 \mathcal{B}_2  \to \eta_L \, , \quad
		Y_2^L + i \ln u_2 \mathcal{B}_2 \to Y_2^L
	\end{aligned}
\end{align}
and similarly for $\eta_R,Y_2^R$. 
The operator $V_{1}$ becomes
\begin{align}
	\begin{aligned}
		V_{1} \to V_{1} &=  e^{ ( \phi^{1}  +  \chi^{1} - \chi^{2} - \frac{1}{\sqrt{n (n+1)}}  \eta ) /b} |e^{i Y_1^L} - e^{i Y_2^L} |^2 \, ,
	\end{aligned}
\end{align}
which can be put back as one of interaction terms.
Notice that from \eqref{kin} there are insertions of 
\begin{align}
e^{-  \psi_2^- \psi_1^+ } = 1 -   \psi_2^- \psi_1^+ = 1 - e^{ - i Y_2^L + i Y_1^L} \, , 
\end{align}
where the vertex operators are inserted.
The vertex operators are changed as
\begin{align}
	V_\nu (z_\nu)  \to  V_\nu (z_\nu) =|\psi_1^+ \psi_2^+| ^2 \left[ \prod_{a=3}^{n } |\mu_a^\nu |^{ 2 ( j_{a}^\nu - j_{a-1}^\nu ) }  \right]
	e^{\sum_{a=3}^{n} (\mu^\nu_a \Gamma_a - \bar \mu^\nu_a \bar \Gamma_a )} e^{b \sum_{a=1}^n j_a ^\nu \phi_a + \phi^{2} /b}
\end{align}
and
\begin{align}
	V^{\chi;\eta}_\nu (z_\nu) \to 	 V^{\chi;\eta}_\nu (z_\nu) = e^{b \sum_{j=1}^{n-1} l_j ^\nu \chi_j - \chi^{2}/b} e^{b \sqrt{n (n+1)} ( ( m ^\nu -   \frac{2}{n (n+1) b^2} )  \eta_L + ( \bar m ^\nu -   \frac{2}{n (n+1) b^2} ) \eta_R )} \, .
\end{align}
The other is the extra insertions of vertex operators 
\begin{align}
	V_b (y_p^2) = e^{ (\phi^1 - \phi^{2}  -  \chi^{1} + \chi^{2}  + \frac{1}{\sqrt{n (n+1)}}  \eta) /b} |e^{- i Y_2^L}|^2 
\end{align}
 for $p=1,2,\ldots , N-2 - S_2$.

Similarly, we integrate out $\beta_a,\gamma_a$ for $a=3,4,\ldots,n$.
The vertex operators are now
\begin{align}
	\begin{aligned}
	&	\Psi_\nu (z_\nu)  \to 	\Psi_\nu (z_\nu)   = \prod_{l=1}^n  |\psi_l^+|^2  \\
	& \quad \times e^{b \sum_{a=1}^n j_a ^\nu \phi_a + \phi^{n} /b  + b  \sum_{j=1}^{n-1} l_j ^\nu \chi_j + b \sqrt{n (n+1)} ( ( m ^\nu - \frac{1}{(n+1) b^2} )  \eta_L + ( \bar m ^\nu - \frac{1}{(n+1) b^2} ) \eta_R )} \, .
    \end{aligned}
\end{align}
The interaction terms are 
\begin{align}
	\begin{aligned}
		V_{l} = e^{ (  - \phi^{l-1} +   \phi^{l}  +  \chi^{l} - \chi^{l+1}  - \frac{1}{\sqrt{n(n+1)} } \eta)/b} |\psi_l^+-  \psi_{l+1}^+|^2  
	\end{aligned}
\end{align}
for $l=2,3,\ldots , n-2$ and
\begin{align}
	V'_{l} = e^{( \phi^{l-1} - \phi^{l} -   \chi^{l-1} + \chi^{l}  + \frac{1}{\sqrt{n(n+1)} }\eta)/b} |\psi_l^-|^2 
\end{align}
for $l=2,3,\ldots , n-1$. Moreover, we have
\begin{align}
	\begin{aligned}
		&V_{1} = e^{ ( \phi^1  +  \chi^{1} - \chi^2  - \frac{1}{\sqrt{n(n+1)} } \eta ) /b} |\psi_1^+ - \psi_2^+ |^2  \, , \\
		&V_{n-1} = e^{ ( - \phi^{n-2} + \phi^{n-1}  + \chi^{n-1}  - \frac{1}{\sqrt{n(n+1)} } \eta) /b}  |\psi_{n-1}^+ - \psi_n^+|^2  \, , \\
		&V_n = e^{\phi_n/b} 
	\end{aligned}
\end{align} 
and
\begin{align}
	\begin{aligned}
		&	V'_{1} = e^{ ( - \phi^1  +  \chi^{1}  + \frac{1}{\sqrt{n(n+1)}} \eta)/b} |\psi_1^-|^2 \, , \\
		&	V'_n = e^{ ( \phi^{n-1}- \phi^{n}  - \chi^{n-1} + \frac{1}{\sqrt{n (n+1)}}  \eta)/b}  |\psi_n^-|^2  \, .
	\end{aligned}
\end{align}
For $V_n$, we have performed a self-duality of Liouville field theory.
The kinetic terms are similar to those of \eqref{scosetaction}.
Only differences are no $(\beta_i , \gamma_i)$ now and the shifts of background charges for $\phi^n, \eta , Y_i$ as
\begin{align}
	Q_{\phi^n} = b + b^{-1} \, , \quad Q_\eta = \sqrt{\frac{ n}{n+1}} b^{-1} \, , \quad Q_{Y} = i \, .
\end{align}

\subsection{Structure of the dual theory}

We have written down correlation functions of the super coset \eqref{KScoset}  in terms of a different theory.
In this subsection, we show that the dual theory is indeed $\mathfrak{sl}(n|n+1)$ Toda field theory in \cite{Ito:1990ac,Ito:1991wb}.

As obtained in the previous section, we have two types of interaction terms $V_l,V'_l$ with $l=1,2,\ldots , n$.
We split the interaction terms $V_l$ (more precisely speaking the corresponding screening operators) into two parts as
\begin{align}
V_{l} = V_{l,1} - V_{l,2} 
\end{align}
with
\begin{align}
	V_{l,1} = e^{ (  - \phi^{l-1} +   \phi^{l}  +  \chi^{l} - \chi^{l+1}  - \frac{1}{\sqrt{n(n+1)} } \eta)/b} \psi_{l+1}^+ \, , \quad 	V_{l,2} = e^{ (  - \phi^{l-1} +   \phi^{l}  +  \chi^{l} - \chi^{l+1}  - \frac{1}{\sqrt{n(n+1)} } \eta)/b} \psi_{l}^+
\end{align}
for $l=1,2,\ldots , n-1$. Here we have set $\phi^0 = \chi^n = 0$.
We can check that the Gran matrix for $V_l '$ and $V_{l,1}$ are the same as that in the bosonic case.
This implies that there is a special transformation of fields such that the fermions (or $Y_l$) are removed from the set of screening operators. It was shown that this is indeed the case for the fermionic FZZ-duality in \cite{Creutzig:2010bt}. After the transformation, we can apply the same reflection relations and change of variables as in the bosonic case. We then obtain screening operators such as
\begin{align}
	V'_{l} = e^{( - \phi^{l-1} + \phi^{l} +  \chi^{l-1} - \chi^{l}  + \frac{1}{\sqrt{n(n+1)} }\eta)/b}  \psi_l^- \, , \quad 
	V_{l,1} = e^{ ( \phi^{l} -   \phi^{l+1}  - \chi^{l-1}  +  \chi^{l}  - \frac{1}{\sqrt{n(n+1)} } \eta)/b} \psi_{l}^+ 
\end{align}
with $\phi^0 = \chi^0 = \phi^{n+1} = \chi^n = 0$.
For the other set of screening operators $V_{l,2}$, the Gram matrix with $V_l '$ are different from that for the bosonic case. This implies that the fermions  (or $Y_l$)  cannot be decoupled by a transformation of fields.
Even so, we can perform the same reflection relations and change of variables as in the bosonic case, and the final result turns out to be quite simple as
\begin{align}
	V_{l,2} = e^{ ( \phi^{l} -   \phi^{l+1}  - \chi^{l-1}  +  \chi^{l}  - \frac{1}{\sqrt{n(n+1)} } \eta)/b} \psi_{l+1}^+ 
\end{align}
for $l=1,2,\ldots,n-1$ and $V_{n,2} = 0$. 
Indeed, $V_l '$ and $V_l = V_{l,1} - V_{l,2}$ are the fermionic screening operators obtained in \cite{Ito:1990ac,Ito:1991wb}.

For the vertex operator, we again perform the same reflection relations and change of variables as in the bosonic case. The vertex operators become
\begin{align}
	 	\Psi_\nu (z_\nu)   =  e^{b ( \sum_{a=1}^n j_a^\nu  \phi_a  +  \sum_{j=1}^{n-1} l_j ^\nu  \chi_j + \sqrt{n (n+1)} (  m ^\nu   \eta_L + \bar m ^\nu \eta_R ))}
\end{align}
as one may have expected.

\section{Conclusion and discussions}
\label{sec:conclusion}

In this paper, we derived correlator correspondences among two dimensional conformal field theories with W-algebra symmetry. Combined with the matchings of symmetry algebra, we can thus show the equivalences of dual theories. We examined several examples, and the most fundamental one may be the duality between the coset \eqref{Wcoset} and $\mathfrak{sl}(n)$ Toda field theory, which can be regarded as an analytic continuation of coset realization of W$_n$ minimal model proven rather recently in \cite{Arakawa:2018iyk}. Another important example is higher rank FZZ-duality analyzed in \cite{Creutzig:2020cmn}. In this paper, we extended the derivation of the correlator correspondences to all $n$. We also analyzed related coset models and those with additional fermions. 
We examined dualities related to those of VOAs conjectured by Gaiotto-Rap\v{c}\'ak  via brane junction picture in \cite{Gaiotto:2017euk}. We have realized some dualities of VOAs in terms of two dimensional conformal field theory, and it would be important to realize all of them.
The triality of Gaiotto and Rap\v{c}\'ak extends to orthosymplectic groups \cite{Gaiotto:2017euk, Creutzig:2021dda} and this is important since even spin algebras and superalgebras with $\mathcal N=1$ supersymmetry are covered by orthosymplectic cosets. In addition to their appearance as corner VOAs of orthosymplectic gauge theories these coset models also appear as duals to $\mathcal N=1$ higher spin theories \cite{Creutzig:2012ar}. 
There are other related dualities, such as, those constructed by combining the fundamental brane junctions as suggested in \cite{Gaiotto:2017euk,Prochazka:2017qum}, the Fateev's duality in \cite{Fateev:1996ea}, and so on.
They deserve further investigation in the current context as well.

We derived the correlator correspondences by utilizing a first order formulation of coset model, which is a simple way to described the coset algebra like free field realizations of affine Lie algebras. We developed the method by expressing the coset model in the BRST formulation \cite{Gawedzki:1988nj,Karabali:1988au,Karabali:1989dk} and applying the Kugo-Ogima method \cite{Kugo:1979gm}. The result not only reproduces the previous proposal by \cite{Gerasimov:1989mz,Kuwahara:1989xy} but also provides a way to find out proper interaction terms.
In particular, the equivalence between the first order formulation and the GKO construction is kept manifest, which follows the equivalence between the BRST formulation and the GKO construction shown in \cite{Hwang:1993nc}.
The correlator correspondences between the coset \eqref{Wcoset} and $\mathfrak{sl}(n)$ Toda field theory are direct consequence of the first order formulation. On the other hand, the correlator correspondences for higher rank FZZ-duality are more involved and required the reduction methods developed in \cite{Hikida:2007tq,Creutzig:2015hla,Creutzig:2020ffn}.
In this paper, we applied the first order formulation of coset models to realize dualities in two dimensional conformal field theory. However, the formulation itself is a fundamental method to investigate properties of coset models. Therefore, we expect that there should be more applications of the current formulation.

In this paper, we have examined only correlation functions on a Riemann sphere, and it would be interesting to extend the analysis to generic surfaces. It would be not so difficult to examine Riemann surfaces of higher genus, see, e.g., \cite{Hikida:2008pe}. However, it would be rather involved to treat Riemann surfaces with boundaries. The original FZZ-duality on a disk was investigated in \cite{Creutzig:2010bt}, and it would be nice if one could understand dualities involving D-branes in a systematic way.
As mentioned at the beginning of the introduction, W-algebras play important roles in several places of theoretical physics. In fact, one of our prime purposes to initiate this project is to understand the properties of conformal field theories dual to extended higher spin gravities as analyzed in \cite{Creutzig:2018pts,Creutzig:2019qos,Creutzig:2019wfe}. 
In particular, it would be important to incorporate extended supersymmetry to see the relation between superstrings and higher spin gravities, see \cite{Gaberdiel:2013vva,Gaberdiel:2014cha} for $\mathcal{N}=4$ supersymmetry and \cite{Creutzig:2013tja,Creutzig:2014ula} for $\mathcal{N}=3$ supersymmetry. 
The Gaiotto-Rap\v{c}\'ak VOAs are conjectured to be isomorphic to the algebras of \cite{BFM} (see also \cite{Litvinov:2016mgi,Prochazka:2018tlo}), and the algebras were shown in \cite{Rapcak:2018nsl} to be the symmetry of moduli spaces of spiked instantons by Nekrasov. Therefore, the understanding of Gaiotto-Rap\v{c}\'ak  dualities would lead to an extension of Alday-Gaiotto-Tachikawa conjecture \cite{Alday:2009aq,Wyllard:2009hg} relating four dimensional gauge theories and two dimensional conformal field theories with W-algebra symmetry.

\subsection*{Acknowledgements}

We are grateful to T.~Arakawa and T.~Takayanagi for useful discussions.
The work of TC is supported by NSERC Grant Number RES0048511.
The work of YH is supported by JSPS KAKENHI Grant Number 19H01896, 21H04469, 21H05187.

\appendix

\section{Relation among two Bershadsky-Polyakov theories}
\label{sec:BP}

One of the important facts used in \cite{Creutzig:2020ffn,Creutzig:2020cmn} is that there are two types of free field realizations for the same Bershadsky-Polyakov algebra  \cite{Polyakov:1989dm,Bershadsky:1990bg} as found in \cite{Genra1,Genra2}. In particular, we constructed actions corresponding to these free field realizations and proposed  a map between correlation functions evaluated by these two actions. In this appendix, we show that the map can be actually obtained simply by a rotation of fields. It is expected that a similar story holds also for more complicated examples of non-regular W-algebras as analyzed in \cite{Creutzig:2020ffn}, and we would like to return to this important issue in the near future.

For the first realization, we use the action
\begin{align}
	S = \frac{1}{2 \pi} \int d^2 w \left[  \frac{G^{(3)}_{ab}}{2} \partial \phi^a \bar \partial \phi^b - \beta \bar \partial \gamma - \bar \beta \partial \bar \gamma + \frac{\sqrt{g} \mathcal{R}}{4} (Q_1  \phi^1 + Q_2 \phi^2) +\lambda \sum_{l=1}^2 V_l \right] \, .
\end{align}
The background charges for $\phi^a$ are
\begin{align}
	Q_1 = b + 1/b  \, , \quad Q_2 = b  \label{bg}
\end{align}
with
 $b = 1/\sqrt{k-3}$.
It is convenient to formulate $(\beta , \gamma )$-system as
\begin{align}
	\beta = - \partial y_L e^{- x_L + y_L}  \, , \quad \gamma = e^{x_L - y_L}
\end{align}
with $x_L(z) x_L(0) \sim - \ln z $ and $y_L (z) y_L (0) \sim \ln z$. We define $x_R,y_R$ in a similar way and also introduce $x = x_L + x_R , y = y_L + y_R$.
The interaction terms are
\begin{align}
	V_1 = \gamma \bar \gamma e^{b \phi_1} \, , \quad 	V_2 = \beta \bar \beta e^{b \phi_2 } \, .
\end{align}
The theory admits the symmetry of Bershadsky-Polyakov algebra and in particular its $U(1)$-generator is given by
\begin{align}
	H = \frac{1}{3 b} (\partial \phi_1 - \partial \phi_2) + \beta \gamma =  \frac{1}{3 b} (\partial \phi_1 - \partial \phi_2) - \partial x \, .
\end{align}

For the second realization, we use
\begin{align}
	S = \frac{1}{2 \pi} \int d^2 w \left[  \frac{G^{(3)}_{ab}}{2} \partial \phi ' {}^a \bar \partial \phi ' {}^b - \beta ' \bar \partial \gamma ' - \bar \beta '  \partial \bar \gamma  ' + \frac{1}{4} \sqrt{g} \mathcal{R}(Q_1  \phi ' {}^1  + Q_2 \phi ' {}^2  ) + \lambda \sum_{l=1}^2 V'_l \right] 
\end{align}
with the background charges \eqref{bg}.
As above, we formulate $(\beta  ', \gamma ')$-system as
\begin{align}
	\beta '= - \partial y' _L e^{- x '_L + y '_L}  \, , \quad \gamma '= e^{x ' _L - y '_L}
\end{align}
with $x ' _L(z) x '_ L(0) \sim - \ln z $ and $y '_ L (z) y '_ L (0) \sim \ln z$. We define $x ' _  R,y' _R$ in a similar way and also introduce $x ' = x '_L + x' _R , y ' = y'_L + y '_R$.
The interaction terms are
\begin{align}
	V '_1 = e^{b \phi '_1} \, , \quad 	V_2 '= \beta ' \bar \beta ' e^{b \phi_2 ' }  \, .
\end{align}
The theory also admits the symmetry of Bershadsky-Polyakov algebra and its $U(1)$-generator is
\begin{align}
	H ' = \frac{1}{3 b}(\partial \phi '_1 + 2  \partial \phi ' _2) - \beta ' \gamma ' =   \frac{1}{3 b } ( \partial \phi '_1 + 2  \partial \phi ' _2) + \partial x ' \, .
\end{align}

We would like to show that the two descriptions are related by a rotation of fields.
We first require $V_1 = V_1 '$, that is,
\begin{align}
	\phi_1 ' = \phi_1 + ( x -  y ) / b \, . \label{cond1}  
\end{align}
We then require that $V_2 = V_2 ' $ up to a total derivative term. Thus we need
\begin{align}
	y ' = c_1 y + c_2 (b  \phi_2 - x + y) \, , \quad b  \phi_2 - x + y =  b \phi_2 ' - x ' + y ' 
\end{align}
with some coefficients $c_1,c_2$.
Finally, we assign that the $U(1)$-generators are the same, i.e.,
\begin{align}
	H =  \frac{1}{3 b} (\partial \phi_1 - \partial \phi_2) - \partial x =  \frac{1}{3 b} (\partial \phi '_1 + 2  \partial \phi ' _2) + \partial x ' = H ' \, .
\end{align}
A solution is given by
\begin{align}
\begin{aligned}
	&	\phi_1 ' = \phi_1 + (  x -  y ) / b \, , \quad  \phi_2 ' = \phi_2 - 2 (x - y)/ b  \, , 
	\\
	&	y ' = (k-3) x - (k- 2 ) y  -   \phi_2 / b  \, ,  \quad x ' = (k -4) x - (k-3) y - \phi_2 / b  \, .
\end{aligned}
\end{align}
The vertex operators are mapped as
\begin{align}
	\gamma ' {} ^\alpha  \bar 	\gamma ' {} ^{\bar \alpha} e^{b (j_1 \phi '_1 + j_2 \phi '_2) } = \gamma^{j_1 - 2 j_2 - \alpha}  \bar \gamma^{j_1 - 2 j_2 - \bar  \alpha}e^{b (j_1 \phi_1 + j_2 \phi_2) } \, .
\end{align}
In particular, the factor relative to these vertex operators is one.

\section{Duality with a theory of a $\mathfrak{gl}(n|n)$-structure}
\label{sec:slnndualtiy}

In \cite{Creutzig:2020cmn} and sections \ref{sec:HR}, \ref{sec:sHR}, we examined the duality related to the duality of $Y_{0,n+1,n}$-algebras in terms of \cite{Gaiotto:2017euk}. A slightly modified duality can be obtained for $Y_{0,n,n}$-algebra, which involves the coset \eqref{slncoset}. In this appendix, we derive correlator correspondences between the coset and the theory with a $\mathfrak{gl}(n|n)$-structure by almost the same analysis as done in section  \ref{sec:HR}. We examine another duality by introducing additional fermions and derive correlator correspondences  by almost the same analysis as done in section \ref{sec:sHR}.

\subsection{Bosonic duality}

\label{sec:slnn}

In subsection \ref{sec:hrg}, we analyzed the coset \eqref{Wcoset}
and showed that its correlation functions match with those of $\mathfrak{sl}(n)$ Toda field theory.
In the coset, we describe $SL(n)_{-1}$ by $n$ pairs of free fermion $(\psi^+_j , \psi^-_j)$ (with a decoupled $U(1)$).
In this appendix, we consider the coset \eqref{slncoset}.
In this case, we describe $SL(n)_{1}$ by $n$ pairs of ghost system $(\beta_j, \gamma_j )$  (with a decoupled $U(1)$).
We work on the Ramond sector such that the conformal dimensions of $(\beta_j , \gamma_j)$  becomes effectively $(1,0)$, see, e.g., \cite{Creutzig:2020ffn,Creutzig:2021cyl} for related issues.

Following the same logic for the case with \eqref{Wcoset}, we can reduce the correlation function of the coset  \eqref{slncoset} to that of the form
\begin{align}
 \left \langle \prod_{\nu=1}^N \Psi_\nu (z_\nu) \right \rangle \, .
\end{align}
The effective action is
\begin{align}
	\begin{aligned}
 S &= \frac{1}{2 \pi} \int d^2 w \left[ \frac{G^{(n)}_{ab}}{2} \left( \partial \phi^a  \bar \partial \phi^b - \partial \chi^a \bar \partial \chi^b \right) +  \frac{\sqrt{g} \mathcal{R}}{4} \sum_{a=1}^{n-1} (Q_\phi  \phi^a - Q_\chi  \chi^a)\right] \\
 & \quad - \frac{1}{2 \pi} \int d^2 w \left[   \sum_{j=1}^n  ( \beta_j \bar \partial \gamma_j + \bar \beta_j \partial \bar \gamma_j ) - \lambda \sum_{l=1}^{n-1} V_l  \right] \, , 
 \end{aligned} \label{appaction}
\end{align}
where the background charges are
\begin{align}
	Q_\phi = b_{(n)} = \frac{1}{\sqrt{k-n}} \, , \quad Q_\chi =  b_{(n+1)} = \frac{1}{\sqrt{k-n+1}} 
\end{align}
and the interaction terms are
\begin{align}
	V_l = |\beta_{l}  \gamma_{l+1}  |^2 e^{b_{(n)} \phi_l}
\end{align}
for $l=1,2,\ldots , n-1$. The vertex operators are of the form
\begin{align}
	 \Psi_\nu (z_\nu)  = \left[ \prod_{i=1}^{n} | \gamma_i^\nu |^{- 2 (j_i^\nu - j_{i-1}^\nu + l_i^\nu - l_{i-1}^\nu)}   \right]e^{\sum_{a=1}^{n-1} ( b_{(n)} j_a ^\nu \phi_a + b_{(n+1)} l_a ^\nu \chi_a ) }
\end{align}
with $j^\nu_0 = l^\nu_0 = 0$.

The integration over $(\beta_j , \gamma_j)$ can be carried out as in section \ref{sec:HR}.
The interaction terms are 
\begin{align}
	\begin{aligned}
		V_{l} &= e^{( \phi_l - \phi^{l} + \phi^{l+1} ) / b_{(n)}+ ( \chi^{l} - \chi^{l+1}) /  b_{(n+1)  } }\\
		&= e^{- ( \phi^{l-1} -  \phi^{l} ) / b_{(n)} + ( \chi^{l} - \chi^{l+1}) / b_{(n+1)} }  
	\end{aligned}
\end{align}
for $l=2,3,\ldots , n-2$ and
\begin{align}
	V'_{l} = e^{( \phi^{l-1} - \phi^{l} ) / b_{(n)} - (  \chi^{l-1} - \chi^{l}) /  b_{(n+1)} } 
\end{align}
for $l=2,3,\ldots , n-1$. Moreover, we have
\begin{align}
	\begin{aligned}
		&V_{1} = e^{ \phi^1  / b_{(n)} + (\chi^{1} - \chi^2 ) / b_{(n+1) }}  \, , \quad
		&V_{n-1} = e^{- ( \phi^{n-2} - \phi^{n-1} )/b_{(n)} + \chi^{n-1} /  b_{(n+1)}} 
	\end{aligned}
\end{align} 
and
\begin{align}
	\begin{aligned}
		&V'_{1} = e^{  -\phi^1  / b_{(n)} + \chi^{1} / b_{(n+1) } } \, , \quad
		&V'_{n} = e^{\phi^{n-1}/b_{(n)} -\chi^{n-1} / b_{(n+1)}}  \, .
	\end{aligned}
\end{align} 
The kinetic terms are the same as \eqref{appaction} except for no $(\beta_j , \gamma_j)$ now.
The vertex operators are 
\begin{align}
	 \Psi_\nu (z_\nu)  = e^{\sum_{a=1}^{n-1} ( b_{(n)} j_a ^\nu \phi_a +  b_{(n+1)} l_a ^\nu \chi_a) } \, .
\end{align}

We can check that the interaction terms correspond to screening operators for a free field realization of $Y_{n,0,n}[\psi^{-1}]$ with $\psi = -k+n$.
For this, we introduce $\phi^{(1)}_j , \phi^{(3)}_j$ with $j=1,2,\cdots , n$. The normalization is
\begin{align}
	\phi^{(1)}_j (z) \phi^{(1)}_l (0) \sim - \frac{1}{h_2 h_3} \delta_{j,l} \ln z \, , \quad
	\phi^{(3)}_j (z) \phi^{(3)}_l (0) \sim - \frac{1}{h_1 h_2} \delta_{j,l} \ln z 
\end{align} 
with
\begin{align}
	h_1 = i \sqrt{k - n} \, , \quad h_2 = \frac{i}{\sqrt{k - n }} \, , \quad h_3 = - i \frac{k-n+1}{\sqrt{k-n}} \, .
\end{align}
We may consider the free field realization corresponding to the ordering 
\begin{align}
 \phi^{(1)}_1 	\phi_1^{(3)}\phi^{(3)}_2 \cdots \phi^{(3)}_{n-1}  \phi^{(1)}_{n} \phi^{(3)}_{n}  \, .
\end{align}
The screening operators are
\begin{align}
	\begin{aligned}
		&	V '_ l = e^{- h_3 \phi^{(1)}_l + h_1 \phi^{(3)}_l}  \quad (l=1,2,\ldots , n) \, , \\
		&	V_l = e^{- h_1 h^{(3)}_l + h_3 \phi^{(1)}_{l+1} }\quad (l=1,2,\ldots , n-1) \, .
	\end{aligned}
\end{align}
They indeed reproduce those obtained above.

\subsection{Fermionic duality}

\label{sec:sslnn}

We then consider the coset of the form
\begin{align}
	\frac{SL(n)_k \otimes SL(n)_{1} \otimes SL(n)_{-1}}{SL(n)_{k}} \, . 
\end{align}
Compared with the coset \eqref{slncoset}, $n$ pairs of complex fermions $\psi^\pm_j$ with $j=1,2,\ldots ,n$ are added.

As in the previous examples, we can reduce the problem to compute the correlation function
\begin{align}
 \left \langle \prod_{\nu=1}^N \Psi_\nu (z_\nu) \right \rangle 
\end{align}
with the effective action
\begin{align}
	\begin{aligned}
 S &= \frac{1}{2 \pi} \int d^2 w \left[ \frac{G^{(n)}_{ab}}{2} \left( \partial \phi^a  \bar \partial \phi^b - \partial \chi^a \bar \partial \chi^b \right) +  \frac{\sqrt{g} \mathcal{R}}{4} \sum_{a=1}^{n-1} (Q_\phi  \phi^a - Q_\chi  \chi^a)\right] \\
 & \quad + \frac{1}{2 \pi} \int d^2 w \left[   \sum_{j=1}^n  ( - \beta_j \bar \partial \gamma_j - \bar \beta_j \partial \bar \gamma_j + \psi^+_j \bar \partial \psi^-_j + \bar \psi^+_j \partial \bar \psi^-_j) + \lambda \sum_{l=1}^{n-1} V_l  \right] \, .
 \end{aligned}
\end{align}
Here the background charges are
\begin{align}
	Q_\phi =  Q_\chi = b = \frac{1}{\sqrt{k-n}}
\end{align}
and the interaction terms are
\begin{align}
	V_l = |\psi^+_{l}  \psi^-_{l+1}   +\beta_{l}  \gamma_{l+1}  |^2 e^{b \phi_l} 
\end{align}
for $l=1,2,\ldots , n-1$. The vertex operators are of the form
\begin{align}
	 \Psi_\nu (z_\nu)  = \left[ \prod_{i=1}^{n} | \Gamma_i^\nu |^{- 2 (j_i^\nu - j_{i-1}^\nu + l_i^\nu - l_{i-1}^\nu)}   \right]e^{b \sum_{a=1}^{n-1} ( j_a ^\nu \phi_a +  l_a ^\nu \chi_a ) }
\end{align}
with $j_0 = l_0 = 0$ and
\begin{align}
	\Gamma_1 = \gamma_1 \, , \quad 
	\Gamma_i = \gamma_i + \psi^+_{i-1} \psi^-_i  \quad (i=2,3,\ldots ,n) \, .
\end{align}

The integration over $(\beta_a , \gamma_a)$ can be carried out as in section \ref{sec:sHR} along with the change of variables
\begin{align}
\Gamma_i = \gamma_i + \psi^+_{i-1} \psi^+_i \to \gamma_i \, .
\end{align}
The interaction terms are 
\begin{align}
		V_{l}  = e^{ (- \phi^{l-1} + \phi^{l}  +  \chi^{l} - \chi^{l+1}) / b } |\psi^+_l - \psi^+_{l+1}|^2  
\end{align}
for $l=2,3,\ldots , n-2$ and
\begin{align}
	V'_{l} = e^{( \phi^{l-1} - \phi^{l} - \chi^{l-1} + \chi^{l}) /  b } |\psi^-_l| ^2 
\end{align}
for $l=2,3,\ldots , n-1$. Moreover, we have
\begin{align}
	\begin{aligned}
		&V_{1} = e^{ ( \phi^1   + \chi^{1} - \chi^2 ) / b } |\psi^+_1 - \psi^+_2|^2  \, , \quad
		&V_{n-1} = e^{( - \phi^{n-2} + \phi^{n-1} + \chi^{n-1} )/b  } |\psi^+_{n-1} - \psi^+_n| ^2
	\end{aligned}
\end{align} 
and
\begin{align}
	\begin{aligned}
		&V'_{1} = e^{  ( -\phi^1   + \chi^{1} ) /  b } |\psi_1^-|^2  \, , \quad
		&V'_{n} = e^{( \phi^{n-1} -\chi^{n-1} )/  b} |\psi_n^-|^2 \, .
	\end{aligned}
\end{align} 
The vertex operators become
\begin{align}
	 \Psi_\nu (z_\nu)  = e^{b \sum_{a=1}^{n-1} ( j_a ^\nu \phi_a +  l_a ^\nu \chi_a ) } \, .
\end{align}
The theory can be identified with the $\mathfrak{gl}(n|n)$ Toda field theory, see \cite{Creutzig:2012sf}.


\providecommand{\href}[2]{#2}\begingroup\raggedright\endgroup

\end{document}